\documentclass[11pt,a4paper]{article}
\usepackage[utf8]{inputenc}
\usepackage{amsmath, amsfonts, amssymb, mathtools}
\usepackage{apacite, natbib}
\usepackage{graphicx, subcaption, float, pdflscape}
\usepackage{booktabs, multirow, tabularx, longtable, threeparttablex}
\usepackage{url}
\usepackage{fancyhdr, geometry, setspace}
\usepackage{xcolor}
\usepackage[toc,page]{appendix}  
\usepackage{ragged2e}
\usepackage{textcomp}
\usepackage{bbm}
\usepackage[textsize=tiny]{todonotes} 

\usepackage[colorlinks=true,
            linkcolor=blue,
            citecolor=blue,
            urlcolor=blue,
            filecolor=blue]{hyperref}

\newcolumntype{C}[1]{>{\centering\arraybackslash}p{#1}}

\begin{document}
	
\begin{titlepage} 
	\begin{center}		\vspace*{1cm}  
		\textsc{\Large Tracing the Genetic Footprints \\ of the UK National Health Service
        \footnote{\textbf{Corresponding author:} Nicolau Martin-Bassols, {nicolau.martinbassols@unibo.it}.
        We thank 
Silvia Barcellos,
Jonathan P. Beauchamp,
Daniel Belsky,
Leandro Carvalho,
Dalton Conley,
David Evans,
Jason Fletcher,
Titus Galama,
Qiongshi Lu,
Teresa Molina,
David Molitor,
Carol Propper,
Julian Reif,
Hannes Schwandt,
Michael Stepner,
Patrick Turley,
Loic Yengo, and participants at several seminars and conferences 
        for useful discussions and suggestions. 
        This research has been conducted using the UK Biobank Resource under Application Number 74002. The work is also based on data provided through www.visionofbritain.org.uk and uses historical material which is copyright of the Great Britain Historical GIS Project and the University of Portsmouth. The project is funded by the Italian Ministry of Universities and Research (MUR) through PRIN 2022 – PNRR (CUP: J53D23015110001, PI: De Cao). De Cao and von Hinke additionally acknowledge support from the ERC under grant agreement No. 101170376 (HARSH) and 851725 (DONNI) respectively. This research is also supported by the European Union’s Horizon 2020 research and innovation programme under the Marie Sklodowska-Curie grant agreement (ESSGN 101073237). The views expressed in this publication are those of the authors and do not necessarily reflect those of the MUR, European Union, MSCA Horizon Europe, or ESSGN. Neither MUR, the European Union, nor the granting authority or ESSGN can be held responsible for them.
        }}\\[0.5cm]
  		
		Nicolau Martin-Bassols\textsuperscript{1},  
        Pietro Biroli\textsuperscript{1},
        Elisabetta De Cao\textsuperscript{1}, \\
		Massimo Anelli\textsuperscript{2}, 
        Stephanie von Hinke\textsuperscript{3},  
        Silvia Mendolia\textsuperscript{4}
        \\[0.5cm]
        
		\textsuperscript{1}University of Bologna,
        \textsuperscript{2}Bocconi University \\
        \textsuperscript{3}University of Bristol, IFS 
        \textsuperscript{4}University of Turin \\[1cm]

\date{December 2025}


\begin{abstract}

   The establishment of the UK National Health Service (NHS) in July 1948 was one of the most consequential health policy interventions of the twentieth century, providing universal and free access to medical care and substantially expanding maternal and infant health services. In this paper, we estimate the causal effect of the NHS introduction on early-life mortality and we test whether survival is selective. We adopt a regression discontinuity design under local randomization, comparing individuals born just before and just after July 1948. Leveraging newly digitized weekly death records, we document a significant decline in stillbirths and infant mortality following the introduction of the NHS, the latter driven primarily by reductions in deaths from congenital conditions and diarrhea. We then use polygenic indexes (PGIs), fixed at conception, to track changes in population composition, showing that cohorts born at or after the NHS introduction exhibit higher PGIs associated with contextually-adverse traits (e.g., depression, COPD, and preterm birth) and lower PGIs associated with contextually-valued traits (e.g., educational attainment, self-rated health, and pregnancy length), with effect sizes as large as 7.5\% of a standard deviation. These results based on the UK Biobank data are robust to family-based designs and replicate in the English Longitudinal Study of Ageing and the UK Household Longitudinal Study. Effects are strongest in socioeconomically disadvantaged areas and among males. This novel evidence on the existence and magnitude of selective survival highlights how large-scale public policies can leave a persistent imprint on population composition and generate long-term survival biases.


    \vspace{0.5cm}

\textbf{Keywords}: Early-life, Health systems, Survival bias, Infant Mortality, Genetics, Polygenic Index, UK Biobank, ESSGN\\
\textbf{JEL classification}: I10, I38, C21
	\end{abstract}	
	

	\end{center}


\end{titlepage}

\begin{flushright}
\textit{“When under changed conditions of life the habits of an animal change, certain parts ... will be reduced ... and others ... increased.”}\\
\vspace{0.1cm}
\textit{— Charles Darwin, \textit{On the Origin of Species} (1859), Ch.~5: \textit{Laws of Variation}}\\[1cm]
\end{flushright}
	
	\section{Introduction}
	\onehalfspacing

Major expansions of public healthcare access rank among the most consequential social policies a government can implement \citep{wust2022universal}. These reforms establish state responsibility for population health and extend medical, hospital, and preventive services to groups previously excluded from care. Early examples include Bismarck’s introduction of compulsory health insurance in Germany in the 1880s \citep{bauernschuster2020bismarck}, the Nordic maternal and child health programs of the 1930s \citep{bhalotra2017infant,butikofer2019infant}, the establishment of the UK National Health Service (NHS) in 1948 \citep{rivett1998cradle}, the implementation of Medicaid in the US during the 1960s \citep{currie1996health,currie1996saving,goodman2018public}, and more recently the Affordable Care Act in the US \citep{blumenthal2015aca,aboulafia2025aca}. Collectively, these policies transformed health systems in their respective countries, sharply reduced morbidity and mortality—particularly among infants and disadvantaged populations—and laid the foundation for the modern welfare state \citep{wust2022universal}. Through their impact on mortality, these policies may also have altered the composition of surviving cohorts, by disproportionately improving survival among individuals with particular social, economic, or biological characteristics. Such selection could change the observed distribution of traits in later-born populations \citep{nobles2019detecting, currie20169}.

One of the most influential healthcare reforms of the twentieth century was the creation of the UK's NHS. Introduced on July 5, 1948, the NHS established universal, tax-funded healthcare for the entire population and, from one day to the next, extended comprehensive medical coverage to roughly 30 million people in England and Wales who had previously lacked access to essential care \citep{rivett1998cradle}. The establishment of the NHS eliminated most financial barriers to treatment, reorganized hospitals under public ownership, and guaranteed equal access to health services regardless of income. Yet, although this unprecedented reform fundamentally revolutionized health policy in postwar Britain, its health and economic effects have received surprisingly little attention \citep{jeffrey2023essays, luhrmann2018}. Given its scale and profound impact \citep{Taylor1951}, it offers a unique setting to study how large-scale health interventions can alter patterns of survival and mortality, and with that, population composition.

Selective survival occurs when survival probabilities differ systematically across individuals based on socioeconomic, behavioral or biological traits, leading to changes in the composition of a population. 
Identifying such selective effects empirically requires information on individuals’ characteristics measured prior to exposure, which is generally not available. Genetic data provide a unique opportunity in this respect: since genotypes are fixed at conception, any change in the genetic composition of a population following a policy reform can signal selection. Furthermore, genomic and biostatistical techniques such as Genome Wide Association Analysis (GWAS) allow us to map genetic variation onto phenotypic traits, suggesting the potential characteristics of newly surviving (selected) individuals \citep{benjamin2024social}. The use of genetic data in this context can therefore provide novel evidence on how social and policy environments can shape population composition and structure—something that, although theoretically modeled \citep{nobles2019detecting}, has limited empirical evidence \citep{furuya2024separating, zhou2024genetic}. Moreover, it can empirically test for the presence, direction and magnitude of selective survival. These quantifications are crucial for understanding demographic change and for assessing the extent to which the scarring effects of policies and other major events \citep[e.g.,][among others]{currie2014we, almond20061918} can partially reflect selection or survival bias.

In this paper, we examine the effects of the introduction of the NHS on early-life mortality and explore whether these effects led to \textit{selective} survival. Specifically, by linking novel, newly-digitized and highly disaggregated historical data on early-life deaths, we estimate the impact of the NHS on (cause-specific) infant mortality rates (IMR) and stillbirth rates, as well as on polygenic indexes (PGIs)—biostatistical measures capturing individuals’ genetic ``propensity'' toward specific phenotypes. Our empirical strategy follows a regression discontinuity design (RDD) under local randomization, comparing individuals born just before and just after July 1948, who are otherwise similar in observable and unobservable characteristics. We draw on several key data sources. First, we use historical Registrar General’s Statistical Review records, which provide detailed information on (cause-specific) early-life mortality at yearly and weekly levels across geographical units in England and Wales. Second, we utilize key population studies with genotypic information that include cohorts exposed to the NHS in early-life—namely, the UK Biobank (UKB), the English Longitudinal Study of Ageing (ELSA), and the UK Household Longitudinal Study (Understanding Society, USoc). 

Our results are threefold. 
\textit{First}, we find a substantial reduction in early-life mortality coinciding with the introduction of the NHS. Our estimates indicate a 17\% decline in the infant mortality rate, corresponding to a reduction of 6.5 deaths per 1{,}000 live births. To shed more light on the drivers of this decline, we turn to newly digitized, highly disaggregated \textit{weekly} data on causes of death in so-called ``great towns'' of England and Wales. These analyses show reductions in rates of stillbirths (8\%) and in particular in mortality rates from diarrhea under the age of two (34\%).  Complementary evidence from newly digitized annual cause-specific mortality data confirms large declines in diarrhea mortality and in mortality from congenital conditions, alongside smaller but meaningful reductions in premature birth, birth injuries, and haemolytic disorders.
We find no evidence of a sizable change in maternal mortality.

\textit{Second}, we show that the NHS left a genetic footprint on the affected cohorts, presenting evidence of selective survival. 
We leverage individual genetic data from the UKB and use month of birth as the running variable. 
We consistently find that post-NHS cohorts present higher average PGIs associated with contextual adverse conditions, such as depression and chronic obstructive pulmonary disease (COPD), and lower PGIs linked to contextual favorable traits, such as self-rated good health and educational attainment.\footnote{We here use the terms ``contextual adverse'' or ``contextual favorable'' to highlight the fact that the interpretation of traits being ``advantageous'' or ``disadvantageous'' are highly context-dependent. Sickle cell disease is a well-known example, where two copies of the minor allele cause the disease, but one copy provides protection against malaria. Hence, it is advantageous in areas with high malaria exposure, but not in areas without. In other words, PGIs cannot be used to infer genetic ``fitness'', since this depends on one's environment; it is not an absolute measure.} 
For both increases and decreases, effect sizes are as large as 7.5\% of a standard deviation. 
These findings are robust to different model specifications and datasets. 
Specifically, we find similar results when using the nationally representative ELSA and USoc datasets, which, together with the UKB, comprise almost all available genetic data cohorts with individuals old enough to have been born around the NHS cutoff. 
We also rule out the presence of selective fertility by showing that the NHS had no significant impact on crude birth rates, nor on household composition (i.e., the birth order of individuals born around the cutoff and the number of (younger and older) siblings, which indicates their parents' completed fertility).

\textit{Third}, to better characterize the selection and understand who the survivors are, we consider within-family variation as well as heterogeneity by region and gender. 
Firstly, we examine a subsample of siblings in the UKB who were born within a short window before and after the NHS introduction, and estimate the impact of the NHS on their PGIs, while (i) controlling for siblings’ PGIs or (ii) using a family fixed effects design. Holding family characteristics constant, 
 our results remain consistent with selection: the most fragile sibling survives. 
Secondly, we exploit local area–level variation in infant mortality in 1947 (i.e., pre-NHS) as a proxy for social disadvantage and examine whether the impact of the NHS varies systematically between areas with higher versus lower baseline mortality. We find substantially larger effects on the genetic distribution in areas with the highest mortality rates, providing further evidence that the NHS enabled the survival of the most disadvantaged. This is also consistent with historical records, which highlight that-in case of limited capacity-priority was given for medical as well as social or environmental reasons. The latter predominantly included women living in overcrowded or unsanitary conditions. Assuming that such unsuitable grounds were more common in areas with initially high infant mortality rates, this is likely to at least partially drive our estimates. 
Finally, we stratify our results by gender and find that the effects are largest for males, consistent with the male frailty hypothesis \citep{kraemer2000fragile,eriksson2010boys}.

Our work makes four contributions to the literature. 
Firstly, we add to the growing body of research examining the impact of large-scale health policies on population health and socioeconomic outcomes (for a recent review, see \citet[][]{wust2022universal}) by exploring one of the most important healthcare reforms of the twentieth century. Research on the US  shows how the introduction and later expansion of Medicaid led to reductions in nonwhite mortality ranging from 20\% for children below 14, to 31\% among children aged one to four and neonates \citep[e.g.,][]{currie1996health,currie1996saving,goodman2018public}.\footnote{These calculations are performed in \citet{goodman2018public}. Several studies document how the long-run positive impact of Medicaid on health and human capital outcomes \citep[e.g.,][]{wherry2016saving,currie2008has,miller2019long, brown2020medicaid,goodman2021long}.} 
Extensive evidence on high-intensity policies for at-risk populations comes from the Nordic countries, again highlighting important positive impacts on children \citep{wust2022universal}. 
The Well-Child Programs, for example, introduced in the 1930s in Denmark, Norway, and Sweden,  targeted infants within their first year of life, leading to significant improvements in infant survival \citep{wust2012early,hjort2017universal, butikofer2019infant, bhalotra2017infant}.
Similarly, \citet{bauernschuster2020bismarck} investigate the introduction of compulsory health insurance among blue-collar workers in the German Empire in 1884, a pioneering effort toward inclusive health care access. This so-called Bismarck's health insurance also affected wives and children of the insured, leading to a 22\% reduction in mortality until the beginning of the 20th century,  largely due to a drop in infectious diseases.

Although the NHS is the first fully universal public health policy of the 20th century \citep{rivett1998cradle}, little empirical research has explored its effects. 
Research on universal large-scale health policies is indeed methodologically challenging, as it is difficult to have a good control group \citep{wust2022universal}.  To date, only two studies investigate the causal impact of the NHS, but with a focus on long-run outcomes, as well as on the children of those exposed \citep{luhrmann2018, jeffrey2023essays}. Their findings suggest that individuals exposed to the NHS at birth are more likely to obtain a degree or a higher qualification, have better self-reported health in adulthood, are less likely to suffer from long-term limiting illnesses, and have improved age-specific survival rates. 
They provide descriptive evidence of a drop in IMR following the introduction of the NHS, driven by neonatal deaths. 
Our research contributes new \textit{causal} evidence on the contemporaneous effects of the NHS. It shows that the NHS significantly and immediately reduced infant mortality rates, visible from \textit{weekly} mortality data. It provided greater protection for the most vulnerable individuals—particularly those with genetic traits contextually associated with poorer health outcomes and those born into the most disadvantaged areas.

Secondly, we systematically quantify perinatal and antenatal mortality using historical registers as well as genomic data. 
Registers allow us to quantify the immediate impact of the introduction of the NHS on early-life survival, while genomic data allow us to move beyond traditional demographic measures and directly assess selective survival—that is, whether the genetic composition of those who survived the prenatal, birth, and early-life period changed following the introduction of the NHS. 
Only two studies have explored a similar question empirically. 
\citet{furuya2024separating} document a correlation between in utero exposure to higher infant mortality environments and PGIs for educational attainment, but their analysis is observational and does not isolate causal selection. \citet{zhou2024genetic} study exposure to the Dutch Hunger Winter in a sample of 723 individuals and focus on one trait (BMI), finding effects close to zero.
We contribute to this literature by causally identifying the impact of a universal health insurance reform on the genetic composition of survivors, broadening the set of PGIs analyzed to capture a richer spectrum of traits, and providing the first evidence of a causal genetic selection effect of substantial magnitude. With the decreasing cost of genomic data collection \citep{muir2016real}, this approach will become increasingly feasible, enabling more precise assessments of selection processes across a wider range of policy and environmental contexts.

Third, our paper speaks to the literature on the long-term impacts of early-life exposures \citep{currie2015early,almond2018childhood}. When examining the effect of a policy or shock on later outcomes, we are by definition restricting the analysis to individuals for whom those later outcomes are observed. If mortality risk is disproportionately different among certain groups of the population, treatment effects on the survivors will be biased. Identifying and quantifying this survival bias remains an unaddressed limitation \citep{almond2018childhood}. Scholars often discuss its direction or try to grasp its magnitude making various assumptions \citep{nobles2019detecting}. For example, \citet{black2007cradle} examine the impact of birth weight on long-term labor market outcomes using a twin birth design. They highlight the potential for substantial underestimation of the effect by showing that birth weight has a much larger impact on the Apgar score when using the full sample of twins, compared to the sample restricted to twin pairs in which both twins survive. \citet{goodman2021long} investigates the long-term impact of childhood Medicaid eligibility and discusses how the policy may skew the composition of survivors toward those more likely to have a disability, and bounds the treatment effects to quantify the bias. In this paper, we propose an empirical approach to test for selective survival or mortality by exploring changes in genetic profiles before and after the implementation of an early-life policy or the experience of a major shock. 

Finally, we expand the literature on social-science genomics by illustrating how PGIs can be leveraged as outcomes, not only as predetermined covariates, in the context of quasi-experimental designs. 
A growing number of studies have used PGIs to investigate the genetic architecture of socioeconomic outcomes and their interaction with environmental exposures \cite[for a review, see][]{Mills2020review,biroli2022economics}. 
While most applications treat PGIs as explanatory variables included on the right-hand side of regression models, we adopt a novel approach by using PGIs as outcomes. 
We exploit the fact that genetic endowments are fixed at conception to trace the composition of the potential full birth cohort prior to survival-related selection, and use this to test whether shocks—such as the introduction of the UK NHS—alter the genetic composition of the surviving population.
In doing so, we provide empirical evidence on a mechanism that has been theorized but remains difficult to quantify in practice \cite{nobles2019detecting}. 


The rest of the paper is organized as follows. 
The next section describes the NHS and the healthcare system in the UK prior to its implementation. We then provide a brief introduction to genetics terminology necessary to understand our empirical approach, 
Sections \ref{sec:data} and \ref{sec:methods} respectively detail the data and the methodology used in the analysis. 
The main results on the effect of the introduction of the NHS are shown in Section \ref{sec:IMR}, analyzing mortality, and Section \ref{sec:surv}, analyzing selective survival through the lens of PGIs.
Section \ref{sec:selective_survival} investigates potential mechanisms and characterizes the most affected individuals. 
Finally, the paper concludes in Section \ref{sec:conclusion} with a discussion of the findings and their implications for both research and policy.

	
	\section{Institutional Background and Genetic Measures} \label{sec:background}

\subsection{History of the NHS}
Before the establishment of the NHS in 1948, healthcare in the UK was largely private, requiring patients to pay for medical services, with limited free care available through various fragmented systems.
Voluntary hospitals, funded by charitable donations, provided some treatment, while local authority hospitals, remnants of the Poor Law system, catered mainly to the indigent.\footnote{The Poor Law system provided means-tested relief, including medical care for the indigent, through locally administered institutions such as workhouses and infirmaries.} 
The National Insurance Act of 1911 created a system of approximately 6,000 Approved Societies that offered basic and limited healthcare to insured workers in exchange for small deductions from weekly wages, covering roughly 25–33\% of the population. However, this system excluded hospital treatment, medications, and care for dependents, leaving large segments of the population without adequate medical support \citep{rivett1998cradle}, in particular women and children.
Subsequent reforms partially expanded access to healthcare. The 1944 Education Act introduced free medical services for schoolchildren, while public health initiatives—such as tuberculosis sanatoria and vaccination programs—sought to control infectious diseases.
Despite these measures, virtually all pregnant women and pre-school children remained uncovered by either of these free medical services. Social investigations conducted during the inter-war period documented the severe consequences of poverty, poor health, and limited institutional support, with working-class women bearing the greatest burden \citep{webster2002national}.

On July 5 1948, the UK government introduced the National Health Service (NHS), revolutionizing healthcare by making it universally accessible and free at the point of use, funded through general taxation.\footnote{In Northern Ireland, the NHS was implemented under separate legislation: the Health Services Act. In Scotland, the NHS was launched with a separate but very similar act (The NHS Scotland Act) in July 1948, and the responsibility of the services was attributed to the Secretary of State for Scotland \citep{Stewart2003}.  } 
Built on three core principles---free provision of healthcare, access based on clinical need rather than financial means, and the equalization of medical services---the NHS aimed to eliminate disparities in care. 
The main purpose of the National Health Service Act was to provide medical care and advice to everyone in need, including a wide range of services such as those offered by general practitioners, nurses, and hospitals. 
For administrative reasons, these services were organized into three branches: hospital services, local health services, and general practitioner (GP)  services \citep{Taylor1951}.  

In England and Wales, the response was immediate and overwhelming: by July 6, 84\% of the population had already registered with a doctor, rising to 91\% by the end of the month. 
By the close of 1948, an astonishing 96\% of  people were enlisted with a GP, marking a swift and widespread embrace of the new system \citep{rivett1998cradle}.

The important supply-side changes were more gradual.
The NHS was launched without any significant increase in hospitals or medical staff---no new doctors or nurses were added.  
What fundamentally changed was the system through which people accessed and paid for care. 
In England and Wales, the NHS assumed control of almost 3,000 hospitals, including both voluntary and municipal facilities, totaling nearly 388,000 beds \citep{rivett1998cradle}. 
Regional Hospital Boards (RHBs) were established to oversee the organization of care within defined ‘‘natural hospital regions’’. 
 As noted in \citet{digby1998continuity}, the reorganization of hospitals in a centralized structure was one of the most revolutionary elements of the NHS Act.

GPs transitioned to independent contractors, receiving fixed fees per treatment as determined by Executive Councils. 
By September 1948, over 18,000 of the 21,000 GPs had enrolled as NHS contractors. 
These structural changes facilitated an immediate surge in healthcare utilization, with a 13\% increase in consultations and a doubling of prescriptions for proprietary medicines within three months, reflecting the rapid expansion of access to medical services \citep{rivett1998cradle}. 

Doctors reported a rise in the number of women and children seeking medical care, as dependents—previously excluded from the health insurance system—began to access regular health services \citep{digby1998continuity}. Indeed, \citet{Davis2012} highlights the establishment of the NHS as a pivotal moment for maternity services, renewing attention to maternal health, with women and children being among its primary beneficiaries.
Although major hospital construction remained limited until 1955, with funding restricted mostly to upgrades of specific departments and essential infrastructure, there was a focus on improving and reorganizing maternity services, such as through the creation of a new NHS Obstetric List. This identified a limited number of qualified practitioners permitted to look after mothers and babies during the so-called ``confinement period'' \citep[i.e., before birth and during the hospital stay;][]{Taylor1951}. The inception of the NHS certainly advanced maternity and childbirth care, even though these services remained variable for decades. 

Given the distinctive institutional features of the NHS in Northern Ireland and Scotland, as well as key administrative differences relative to the rest of the UK \citep{webster2002national}, we restrict our analysis to England and Wales.\footnote{For example, Scotland integrated teaching hospitals into the regional health structure from the outset, while England and Wales initially proposed exempting them from state control and later nationalized them with separate governance \citep{webster2002national}.}

\subsection{NHS Impact on Health at Birth: Potential Mechanisms}\label{sec:IMRmechanism}

The introduction of the NHS allowed pregnant mothers to deliver in a hospital or receive medical assistance at home, as well as access prenatal and postnatal care, all free of charge. 
While there is limited evidence regarding the proportion of mothers directly affected by this policy, female medical consultations increased after the introduction of the NHS \citep{digby1998continuity, jeffrey2023essays}. 

Health improvements for children affected by the policy at the time of birth could be driven by two main channels. 
First, infants born in hospital settings experience lower rates of infections, reduced complications such as neonatal hypoxia, and improved outcomes for preterm births due to improved neonatal care \citep[see][for evidence from the United States]{backes2020maternal}. Children are also more likely to survive home births if they are medically supervised, for example, with the correct use of forceps in cases of breech presentation, shoulder dystocia, or other complications.
At the same time, maternal mortality is generally lower in hospital-based childbirth, as access to skilled medical professionals, sterile environments, and emergency interventions greatly improves survival rates \citep[e.g.,][]{symonds2023risk}. A reduction in maternal mortality enhances infant survival and supports children’s development throughout early-life.
Second, the introduction of universal healthcare mitigates the financial burden on families, reducing out-of-pocket expenses related to childbirth and allowing for broader access to essential maternal and infant care services, which in turn supports long-term economic stability and well-being \citep{bufe2021financial,Finkelstein2012}.

In our analysis, we are not able to isolate the effect of these separate channels. Nevertheless, all these mechanisms could have contributed to improving individuals' health, and the bundled impact of these channels is the policy-relevant one.

	\subsection{Genetic variants, GWAS, and PGIs}

Human DNA consists of approximately three billion nucleotide pairs arranged in 23 pairs of chromosomes, one inherited from each parent. 
Most nucleotides in human DNA (about 99.9\%) are identical across individuals. 
The remaining variation occurs at specific locations known as polymorphisms. 
The most commonly measured form of genetic variants are single nucleotide polymorphisms (SNPs), which are one-base differences at a given genomic locus. 
SNPs can occur in coding or non-coding regions of the genome and may influence gene expression or protein function, thereby contributing to phenotypic differences across individuals \citep{alberts2015essential}.

Since genetic variation is fixed at conception and measured consistently across individuals, space, and time, it provides a fully predetermined measure that is well suited for studying selection.

The relationship between genetic variants and complex traits has been studied using Genome-Wide Association Studies (GWAS): a powerful, hypothesis-free study design that scans the entire genome to identify genetic variants associated with the outcome of interest. By analyzing large samples and often performing meta-analysis to increase statistical power, GWASs yield statistically precise and replicable associations. These associations are predictive but they should not be interpreted as providing causal or biological mechanistic insights.

Almost two decades of GWASs has highlighted that most traits relevant to social scientists have a highly polygenic genetic architecture \citep{Abdellaoui2023}. Rather than having a ``gene for'' a certain outcome, genetic predispositions reflect the aggregation of many variants, each contributing marginally. As a result, empirical analyses typically rely on polygenic indexes (PGIs), which combine information from a large number of SNPs into a single summary measure.
PGIs are the best linear genetic predictor of an outcome \citep{Mills2020book,beck2021genetic} and are constructed by aggregating the weighted average of effect sizes derived from one or more GWASs. In these indexes, larger effect variants contribute more significantly, reflecting the cumulative genetic contribution to specific phenotypes. 
An individual’s genotype at a given locus, denoted as \(G_{ij}\), can take values of 0, 1, or 2, representing the number of minor alleles they carry (e.g., 0 for no copies, 1 for one copy, and 2 for two copies). The formula for calculating a PGI is represented as:  

\[
PGI_i = \sum_{j=1}^{n} W_j G_{ij}
\]

where \(W_j\) represents the GWAS coefficient of the \(j\)-th variant's association with a particular outcome, and \(G_{ij}\) denotes the genotype of the \(i\)-th individual at the \(j\)-th locus \citep{benjamin2024social, biroli2022economics}.

PGIs are powerful linear predictors, but should be interpreted with care. 
Their predictive performance is inherently tied to the environmental and demographic context of the GWAS discovery sample, conducted primarily in affluent and European-ancestry populations \citep{Martin2019}; their portability to populations with different social or ancestral backgrounds is limited \citep{benjamin2024social}. 
More generally, associations between PGIs and outcomes should not be interpreted as fixed biological relationships: estimated effects reflect gene–environment interplay and may vary across contexts \citep[e.g.,][]{mostafavi2020variable, biroli2022economics}.


	\section{Data} \label{sec:data}
\subsection{The General Register Office}

The General Register Office (GRO), established in 1836 under the \textit{Births and Deaths Registration Act}, was responsible for creating a unified system of civil registration in England and Wales and for compiling official statistics on births, deaths, and marriages. Data published by the GRO were  derived from the civil registration process. When a death occurred, a medical practitioner who attended the deceased issued a medical certificate of cause of death, which was verified and registered locally. Unattended or unexplained deaths were referred to a coroner, whose verdict was then entered in the register. Births were reported by parents or informants within a statutory period. 
These local records were transmitted to the General Register Office, where they were aggregated, tabulated, and published to provide a comprehensive and continuous record of demographic and public health trends \citep{london1952registrar, carrier1960registrar}.

The data collected by the GRO capture information on population health, mortality, births, and other demographic trends. They provide insights into the distribution, size, and characteristics of the population, including information on marriage, mortality, and disease rates. It was a key resource for understanding social and health trends in the UK during its period of publication, serving as one of the principal means through which vital statistics were analyzed and disseminated. 

We digitize data from three complementary sources produced by the Registrar General’s Office.
First, we use district-year-level infant mortality rates (IMR) from the \textit{Registrar General’s Statistical Review}, obtained from the Great Britain Historical Database \citep{GBHD2020}, systematically checked and quality-controlled in \citet{baker2024gxe}. The data cover 1,472 local government districts (henceforth: districts) in England and Wales. The IMR is defined as the number of deaths in the first year of life per 1,000 live births in district $d$ at time $t$:

\[
\text{IMR}_{dt} = \frac{\text{Deaths of infants under one year}_{dt}}{\text{Live births}_{dt}} \times 1000
\]

This measure provides a consistent indicator of the early-life health environment across districts and years.  
For our empirical model that uses annual data, we focus on the years  1946 to 1950 which lie around the NHS introduction, reaching a sample that includes 1,472 district-year units, yielding 2,924 observations before and 2,940 after the NHS implementation.\footnote{The panel is unbalanced due to missing data for some districts and years.}

Second, we collect and digitize novel data from the \textit{Registrar General’s Weekly Returns}, which report weekly births, deaths, and cause of death counts for county boroughs and so-called ``Great Towns'' (i.e., those with a population exceeding 50,000 residents), as well as for Greater London. These weekly publications provide information on infant mortality, stillbirths, and few cause-specific mortality. The latter are specific to infectious diseases and cover all ages, though some are predominantly found in early childhood (e.g., diphtheria, whooping cough and diarrhea, where only the latter is specific to children under two years of age). Following the same approach as with the annual data, we construct corresponding mortality rates per live births to ensure comparability across locations and over time. For our main regressions with weekly data, which focus on the 52 weeks before and after the introduction of the NHS, the sample includes 113 area-year units, yielding 5,876 observations before and 5,871 after NHS implementation.

Third, the \textit{Registrar General’s Statistical Reviews} also provide, for the same urban geographic units, annual summaries with more detailed information on cause of death. We collect and digitize these data, with our empirical analysis focusing on causes of death that are specific to pregnancy and early-life, including Hemolytic Disease of the Newborn, Birth Injury, Congenital Anomalies, Other Infant Diseases, Premature Birth, and Diarrhea under age two. We also consider deaths attributable to maternal mortality, including deaths due to Childbirth-Related Infections, Overall Maternal Mortality, Maternal Sepsis, Abortion-Related Causes (both septic and non-septic), and Maternal Complications such as hemorrhage or embolism. 
This disaggregation allows for a more precise examination of the composition of early-life and maternal mortality and how specific causes evolved over time, particularly in relation to greater access to medical care following the introduction of the NHS. For our main regressions that are based on annual data, we focus on the two years before and after the NHS introduction, the sample includes 145 area-year units, yielding 290 observations before and 290 after NHS implementation.

\subsection{The 1951 Census of Population}
\label{sec:Census}

The 1951 Census of Population was the first post-war census in England and Wales, conducted on April 8, 1951. It collected detailed information on the demographic, economic, and social characteristics of the population, covering housing, employment, education, and family structure. A distinctive feature of the 1951 Census is the use of the Registrar General's Social Class schema, which classified occupations into six groups (I–V, where III covers two sub-groups) based on skill and professional status. This scheme provides a widely used measure of SES in mid-20$^{th}$ century Britain.

Specifically, Social Class I and II correspond to professional and managerial occupations, while Social Class IV and V capture partly skilled and unskilled workers. Class III, which includes both manual and non-manual skilled workers, represents the bulk of the labor force. For our purpose, we use this occupationally based SES classification to construct indicators of high and low SES. These measures will be exploited in our heterogeneity analysis, allowing us to examine whether the effects of the NHS reform varied across the socioeconomic distribution.


\subsection{UK Biobank}\label{sect:UKB}

The UKB is a large-scale biomedical database and research resource containing in-depth genetic, health, and lifestyle information from approximately 500,000 UK participants aged 40 to 69 at recruitment (2006 to 2010). It integrates genetic data, physical and cognitive measures, biochemical assays, and extensive longitudinal health records. 
The UKB recruited participants through 22 assessment centers across the UK. Approximately 9.2 million invitations were sent, yielding a response rate of around 5\%. This low response rate resulted in a sample that is not nationally representative, as it is biased toward individuals with higher education, higher SES and better overall health. Consequently, while UKB provides valuable insights into genetic and health-related research, findings must be interpreted with caution \citep{swanson2012uk}. Recent efforts have aimed to improve national representativeness by creating sample weights based on the UK census \citep{van2024reweighting}.

The UKB has information on the month of birth of individuals, which allows us to identify if individuals were born before or after the implementation of the NHS. It also collects self-assessed health (rated on a scale from one to six, with higher values indicating worse health status), and all participants underwent a nurse-led examination, among others collecting data on their BMI. 

Particularly relevant for our paper is the availability of individual genomic data, derived from blood samples.\footnote{The UKB collected blood samples from approximately 500,000 participants, and genotyping was performed using the UKB Axiom Array, designed to capture genome-wide genetic variation, including SNPs and short insertions and deletions (indels). DNA was extracted from the stored blood samples for this genotyping process. 
} 
%
%
%
%
%
We use PGIs from the Polygenic Index Repository \citep{becker2021resource}.
A challenge in our analysis is deciding which PGIs to examine. Testing all PGIs would require running many regressions, complicating interpretation and increasing the risk of false positives due to multiple hypothesis testing \citep{benjamini1995controlling, dunn1961multiple}. On the other hand, selecting only specific PGIs could introduce publication bias if decisions are influenced by significant results \citep{sterling1959publication}. To mitigate this, 
we focus on PGIs classified into meaningful categories that can be interpreted as capturing some form of resilience in early-life, namely, health, mental health, anthropometric measures, cognition and education \citep{becker2021resource}.  
As a robustness check, we present results using all available PGIs from the repository and adjusting for multiple testing. 

The initial UKB sample consists of 502,336 individuals. We make few needed restrictions. First,  we only keep individuals included in the first PGI repository. This includes all siblings, which we need for the within-family design.\footnote{When PGI weights are derived from the same dataset as that used to construct the GWAS weights, overfitting can introduce bias. To address this, \citet{becker2021resource} construct UKB PGIs using three independent repositories: the UKB is split into three equally sized subsamples, and for each subsample, summary statistics from the other two are used to generate independent PGI weights, avoiding overfitting.
As a result, our estimating sample consists of one-third of the UKB sample that includes siblings.}  This reduces our sample to 148,593 individuals. After narrowing the sample to the relevant time window of 12 (or 24) months before and after the introduction of the NHS, and restricting to individuals born in England and Wales with non-missing month of birth and genotypic information, the sample shrinks to 13,641 individuals. Importantly, and as is common practice in the literature \citep{benjamin2024social}, this sample is restricted to individuals of European ancestry, primarily classified as ``White or Caucasian''. This restriction helps reduce bias from population stratification (systematic differences in genetic ancestry that can confound associations between genetic variants and outcomes), ensuring greater comparability with the GWAS summary statistics used to construct PGIs.

\subsection{English Longitudinal Study of Ageing} \label{sec:ELSA}

The ELSA is a large, ongoing study that collects data on the health, economic, and social circumstances of over 18,000 individuals aged 50 and older in England. The study, which began in 2002, follows participants every other year to examine determinants of aging, including physical and mental health, income, employment, and quality of life. 
Crucially, ELSA is nationally representative of the English population. 

The restricted version of ELSA includes information on the month of birth of individuals, which we use to identify those born before or after the implementation of the NHS. Furthermore, genome-wide genotyping was conducted, and PGIs were constructed using externally valid SNP weights  \citep{ajnakina2022english}. 
Aiming to maximize the overlap with the PGIs available in the UKB, we select PGIs across the categories of socioeconomic traits, mental health and well-being, physical health, anthropometric traits, and intelligence. 
When multiple PGIs exist for the same phenotype from different GWAS, we use the PGI derived from the most recent study.

The initial ELSA sample consists of 21,344 individuals. After restricting the sample to a one-year bandwidth around NHS implementation and to those with genotypic information, the final sample comprises 705 individuals. This sample is further restricted to individuals of European ancestry, leading to a final sample size of 465 individuals.

\subsection{The UK Household Longitudinal Study: Understanding Society
} \label{sec:USoc}

USoc is a large, nationally representative longitudinal study that follows individuals and households in the UK to collect detailed information on their economic, social, and health circumstances. The study began in 2009 and now tracks over 40,000 households across multiple waves. The survey covers a wide range of domains, including income, education, employment, family dynamics, health, and well-being, making it a rich resource for studying intergenerational processes and life-course trajectories. 

For our purposes, we use the restricted-access version of USoc, which includes information on respondents’ quarter and year of birth, allowing us to identify individuals born before and after the introduction of the NHS. In addition, genome-wide genotyping was conducted on a subset of participants of European ancestry, enabling the construction of PGIs. 
USoc does not include pre-constructed PGIs, but using the full genotypic information we generate PGIs following \citet{becker2021resource}.

The original USoc sample consists of over 100,000 individuals. However, the special-license sample that we use, which includes genotypic information, contains only 9,920 individuals. For our main specification, we further restrict the sample to a symmetric one-year bandwidth around the introduction of the NHS, which reduces the analytical sample to 414 individuals.


\section{Methodology} \label{sec:methods}
\subsection{Main empirical strategy}\label{sec:method_rdd}

The NHS was introduced on 5 July 1948. 
Our identification strategy exploits this policy threshold using a regression discontinuity design (RDD) under the \emph{local randomization} framework \citep{cattaneo2022regression}. 
The core assumption is that, within a \emph{local window} around the cutoff, assignment to NHS exposure is as good as random, so that individuals born just before and just after the introduction are comparable in both observed and unobserved characteristics.

In our setting, this interpretation is plausible because pregnancies reaching term around July 1948 were conceived well before the policy was announced and then implemented and, given limited scope for precise timing of delivery with Cesarean sections being rare, treatment status was not manipulable at the margin. 
Importantly, for our genetic analysis, genotypes are fixed at conception and remain invariant over the life course. Therefore, the NHS cannot affect genetic endowments directly; any discontinuity in the distribution of PGIs across cohorts can only arise through selective survival into the observed sample.

Let $S_i$ denote the running variable and $c$ the cutoff date corresponding to the NHS introduction (5 July 1948). Treatment assignment is given by:
\begin{equation}\nonumber
NHS_i = \mathbf{1}(S_i \geq c),
\end{equation}
where $NHS_i=1$ indicates being born at or after the introduction of the NHS. 
Under the local randomization framework of \citet{cattaneo2022regression}, we assume that within a chosen local window $\mathcal{W}$ around the cutoff,
\begin{equation}\nonumber
\left(NHS_i \;\perp\!\!\!\perp\; Y_i(0),\, Y_i(1) \right) \,\big|\, S_i \in \mathcal{W},
\end{equation}
so that treatment is as good as randomly assigned, and $Y_i(0)$ and $Y_i(1)$ denote the potential outcomes without and with NHS exposure, respectively. 
Within $\mathcal{W}$ we estimate the average effect using a simple difference-in-means regression:
\begin{equation} \label{eq:main_OLS}
Y_i = \alpha + \tau NHS_i + \varepsilon_i, \qquad i \in \mathcal{W},
\end{equation}
where $Y_i$ is the outcome of interest for unit $i$
and $\tau$ captures the average effect of NHS exposure.

In practice, the running variable is discrete (week, month, quarter, or year of birth), depending on the temporal resolution available in the data sources. 
We therefore implement the design within local windows of one year around the introduction of the NHS in July 1948 that are compatible with each dataset.\footnote{Note that World War II ended in May 1945 and we avoid selecting windows that overlap with the war. See section \ref{sec:robustness} for robustness to alternative bandwidths.} 
For outcomes measured at the weekly level (early-life mortality), the cutoff is defined as the week containing July 5, 1948, with a window of $\pm 52$ weeks; 
for outcomes measured at the monthly level (PGIs from the UKB, ELSA, and USoc), the cutoff is July 1948, with a window of $\pm 12$ months; 
and for outcomes measured at the annual level (IMR), the cutoff is 1948, with a window of $\pm 2$ years.\footnote{Since infant mortality rates include deaths occurring within the first \textit{year} of life, our measures of early-life mortality necessarily include some births prior to the introduction of the NHS. As a result, estimated effects may reflect not only exposure from birth onwards, but also access to NHS services during the first weeks and months of life among individuals born shortly before July 1948. The extent to which post-birth exposure contributes to the estimates depends on the temporal aggregation of the data. Over half of all deaths in the first year of life occurred in the first four weeks (55\% in 1947 and 58\% in 1948), with an additional 30–31\% occurring in the subsequent five months, underscoring the heightened vulnerability of the first six months of life \citep[see Table 15,][]{GRO1947, GRO1948}.}

Another important difference across outcomes concerns both the populations and the level at which outcomes are observed. Infant mortality pertains to the full birth cohort, and is measured at a local geographical level. In contrast, PGIs are measured at the individual level and are observed only for individuals who survive and agree to be genotyped. Consequently, discontinuities in PGIs are interpreted as changes in the \emph{composition} of surviving cohorts rather than as treatment effects on genetic traits.

Inference follows the logic of the local randomization framework: we report $p$-values based on permutation tests, and construct confidence intervals under interference \citep{rosenbaum2007interference}.
Results are robust to alternative specifications considering different bandwidths, kernels, and multiple hypothesis testing. They are also robust to alternative empirical strategies such as standard sharp RDD, and global mean-comparison approaches such as linear cohort regressions and two-sample t-tests. Alternative specifications, such as a sharp RDD and multivariate linear regression models, are also explored as robustness checks.\footnote{As robustness checks, we estimate (i) a conventional sharp local–linear RDD following \cite{cattaneo2022regression}, using triangular kernel weighting and bandwidth windows aligned with the temporal aggregation of each dataset, and (ii) a multivariate linear regression model comparing pre- and post-NHS cohorts. Both specifications include the same set of controls — region and time fixed effects for the mortality analysis, and sex, interview date, quarter-of-birth indicators, and the first ten ancestry-specific principal components for the genetic analyses.
}

\subsection{Potential threats to identification}
Potential threats to identification are discussed below, including fertility responses or contemporaneous shocks. We show that these concerns are unlikely to generate discontinuities precisely at the NHS cutoff within the local windows used in the analysis.

\paragraph{Selection into fertility.}\label{fertility}
In principle, the introduction of free hospital childbirth under the NHS could affect fertility decisions, either by changing the likelihood of having a child or by altering the timing of conception. Such responses could generate compositional changes in birth cohorts, complicating the interpretation of our estimates.

In practice, however, strong fertility responses around the implementation date are unlikely. First, the availability of free hospital delivery alone is unlikely to have induced a sizable increase in total fertility. The reform primarily affected the conditions under which births occurred, rather than the broader economic or social determinants of fertility. 
Second, biological constraints sharply limit the scope for timing responses. The NHS was announced in January 1948 and implemented on July 5, 1948, following several months of parliamentary debate during which final approval remained uncertain \citep{webster2002national}. Individuals giving birth around the implementation date were therefore already pregnant at the time of the announcement. 

Taken together, these considerations suggest that selection into fertility is unlikely to drive discontinuities at the policy cutoff. Nevertheless, to assess this possibility directly, we test for effects of the NHS introduction on fertility using two complementary approaches.
First, we derive weekly crude birth rates from the Registrar General’s Weekly Returns. Rates are calculated dividing total number of live births by (annual) population estimates coming from Registrar General’s Statistical Review. We report three estimates based on alternative cutoff definitions. The first uses the baseline cutoff of the week including July 5, 1948 to test whether dynamics in our sample generate fertility changes that become visible at the time of the reform. The second suspends the cutoff by three months to capture potential fertility responses to the NHS announcement in January 1948, corresponding to a typical nine-month gestation period. The third examines whether fertility changed after NHS implementation, shifting the cutoff to be nine months after July 1948.
Figure \ref{fig:fertility1} shows that, regardless of the cutoff chosen or the estimation method, all results yield estimates close to zero, with confidence intervals that cross zero.\footnote{Note that there was only limited availability of contraceptive methods at the time, with significant breakthroughs only occurring in the 1960s when the contraceptive pill became widely accessible \citep{murphy1993contraceptive,decao25}. This may partly explain the lack of fertility postponement that we find.} 

Second, the UKB has information on the number of younger and older siblings for each respondent at the time of the interview. We use this information to construct the birth order for each individual in our window. In Appendix Figure \ref{fig:fertility_smallsample}, we estimate the impact of the NHS on being firstborn, on the number of older siblings, and on the number of younger siblings using three alternative samples. For each outcome, we report three estimates based on the alternative cutoffs as defined above. 
Except for a small but significant negative effect for being the older siblings nine months after the introduction of the NHS, the remaining coefficients are small and indistinguishable from zero.
\footnote{Sample sizes are substantially smaller than in our main estimations because these questions were collected only for a subsample of UKB participants. We cannot investigate the impact of the NHS on parental age at childbearing, since parental age is asked only if the respondent's parents are still alive at the time of the interview leaving us with a negligible (and most likely selective) sample to work with.} 



\paragraph{Selection into participation.}

A further threat to identification is selective participation. More specifically, exposure to the NHS could have had long-term health and economic benefits. Given that 
healthier and wealthier individuals are more likely to participate in research studies—a phenomenon known as healthy volunteer bias \citep{van2024reweighting}--the introduction of the NHS could have increased their likelihood to participate in the UKB, and this could be misinterpreted as selective survival.

To empirically test this, we first construct UKB sample-representative weights following \citet{van2024reweighting}, and compute attrition rates for that weighted sample. Second, we consider an intensive margin of survey participation: subsamples of UKB participants have been invited to follow-up interviews. We therefore examine whether exposure to the NHS at birth affects the probability of individuals participating in the UKB and, subsequently, taking part in the follow-up studies. In Appendix Figure \ref{fig:participation_sample}, we present estimates for these outcomes; their signs are noisy with confidence intervals that widely cross the zero line.\footnote{Sample sizes are slightly smaller than in our main estimations because information on current place of residence is not available for the full sample, which is necessary to construct sample weights. }

\paragraph{Parallel events.}
Finally, the occurrence of events coinciding with the introduction of the NHS could confound its effects. Only very minor events happened around the NHS introduction. For instance, bread rationing was introduced on 21 July 1946 and ended on 24 July 1948, just weeks after the NHS was established. This is unlikely to be an issue given that bread rationing itself did not affect bread (or other food) consumption, nor did it lead to any government savings or control over its distribution  \citep{zweiniger1993bread, vonHinke2025bread}. However, food rationing more generally was still in place during this period, with key events including reductions in bacon, fat, and soap rations in 1945; the introduction of potato rationing in 1947; and strikes that disrupted food imports. Meanwhile, petrol rationing underwent repeated adjustments, with restrictions easing in 1948 and ending in 1950. Clothes rationing concluded in 1949, and the gradual dismantling of food rationing started in the early 1950s and culminated in its complete removal in July 1954 \citep{zweiniger2000austerity, kynaston2008austerity, zweiniger1994rationing}.

To explore whether the introduction of the NHS correlates with food consumption and dietary quality, driven largely by post-war food rationing, we plot data from The National Food Survey for the years 1945–1949. We focus on households' diets, showing trends in the nutritional composition of domestic food consumption (Figure \ref{fig:nutrients}) and the average expenditure by food group  (Figure \ref{fig:food_groups}), obtained from the \citet{ministry1951urban}. These show no large changes in any nutrients, food groups, or spending categories, suggesting no correlation with the introduction of the NHS. While changes in the quality of food could have led to health differences, the vast majority of food items were rationed around the introduction of the NHS, with little variation in quantity or quality across individuals, suggesting this is not a plausible explanation of our findings.


\subsection{Within-Family Analysis} 


PGIs capture biological predispositions but also embed relevant environmental conditions correlated with family genetic background.
In this context, we aim to understand whether the selection we observe following the introduction of the NHS reflects less resilient individuals surviving at higher rates after the implementation of the NHS, or whether it instead reflects differences across families, with some benefiting more from the policy or having better or worse access and uptake.

To disentangle these two channels, we complement the locally randomized RDD with a within-family analysis using inferred sibling relationships in the UKB \citep{amin2022higher}.
Within-family comparisons exploit the fact that genetic variants are randomly allocated across siblings at conception, allowing us to difference out shared family background--including genetic nurture operating through parental genotypes--while isolating selection operating within families \citep{young2022mendelian}.
Comparing siblings allows us to hold constant much of the shared family background and genetic ancestry, narrowing in the biological channel, and to test whether estimated discontinuities in PGIs persist once we account for family-level factors. 

We implement two complementary specifications. First, we control for a proxy of the family genetic pool by including a sibling’s PGI as a covariate. 
Second, we estimate models with family fixed effects, which absorb all time-invariant family-level characteristics, including parental background, the shared environment, and genetic ancestry. 
Given the smaller sibling sample, and to maintain statistical power while keeping the design local to the reform, for this analysis we use a symmetric two-year window around July 1948.

The first specification is:

\begin{equation}\label{eq:OLS_sibpgi}
\textit{PGI}_{i}
= \alpha
+ \beta \,NHS_{i}
+ \gamma \,\textit{PGI}^{\text{sib}}_{i}
+ \theta' X_{i}
+ \epsilon_{i}
\end{equation}
where the outcome is the PGI of individual $i$ born within a two-year bandwidth around the introduction of the NHS in July 1948.
The sibling PGI, $\textit{PGI}^{\text{sib}}_{i}$, may come from a sibling born outside the two-year window and proxies for the shared family genetic background. 
The vector $X_i$ includes quarter-of-birth fixed effects, an indicator for sex, linear trends for date of birth (month-year) and interview date, and the first ten genetic principal components. These principal components summarize major axes of genetic variation and provide a statistical control for ancestry-related differences in allele frequencies, thereby reducing bias from population stratification \citep{price2006principal, hellenthal2014genetic}. Robust standard errors are reported.


By controlling for the sibling PGI, this specification controls for a substantial component of the genetic nurture effects that operate at the household level and shape household environments, captured by $\gamma$.
However, this adjustment is incomplete: family-level influences that are correlated with genetic variation but differ across siblings are not fully absorbed by the sibling PGI. As a result, the estimated coefficient on NHS exposure $\beta$ may reflect a combination of selective survival and family-level responses to the reform that vary within families, such as differences in prenatal investments or care-seeking behavior.


A second, more demanding, model compares sibling pairs born before and after the NHS through the inclusion of family fixed effects ($\mu_f$):  

\begin{equation}\label{eq:OLS_famFE}
\textit{PGI}_{if}
= \alpha
+ \beta \,{NHS}_{if}
+ \delta' X_{if}
+ \mu_f
+ \epsilon_{if}
\end{equation}
where the outcome is the PGI of sibling $i$ in family $f$. The sample includes only sibling pairs in which one sibling was born within two years before the NHS and the other within two years after. 
Including family fixed effects ($\mu_f$) forces us to use only within-family variation in PGI and controls for all time-invariant family-level characteristics--such as parental background, environmental factors, or genetic ancestry. 
This design removes all between-family variation, making the model more conservative and providing a stricter test of whether observed discontinuities in PGIs reflect selective survival rather than between-family compositional differences. 
The control vector $X_{if}$ is defined as above. Standard errors are clustered at the family level.


This approach identifies the difference in PGIs for siblings exposed to the NHS at birth and those not exposed, holding constant all shared family-level factors, including parental genetic background, the household environment, and other time-invariant family characteristics. Hence, $\beta$ captures the causal impact of the introduction of the NHS on genetic selection \textit{within} family.

These two approaches reflect a trade-off between statistical power and the extent to which shared family environments and genetic background are controlled for. The first model relies on a larger sample but accounts for potential parental effects less comprehensively, while the second model considers a substantially smaller sample and more stringently controls controls for shared family-level environments and genetic background. 


\section{The Impact of the NHS  on Early-Life Mortality}\label{sec:IMR}
\subsection{Annual Infant Mortality}

The first contribution of our work is to show the effect of the NHS on IMR, which form the foundation for the subsequent analyses in this paper. Identifying a substantial impact on IMR is essential, as it would indicate meaningful demographic shifts that justify investigating potential genetic selection. 
We start by plotting the raw IMR across cohorts for a 20-year period from 1938 to 1958.

\begin{figure}[htbp]
	\centering
	\includegraphics[width=1\textwidth]{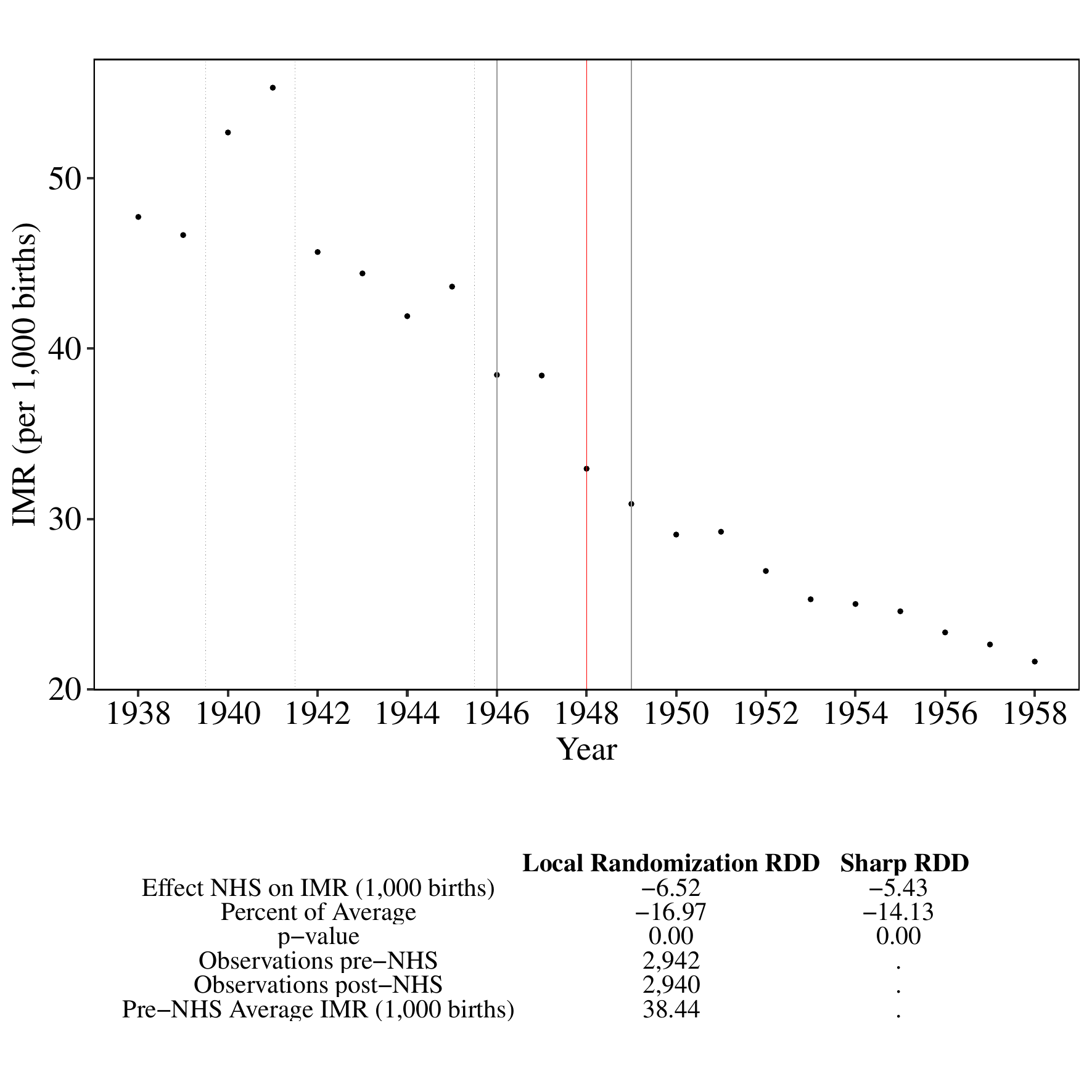} 
	\caption{Effects of the NHS on Infant Mortality Rates}
	\label{fig:2}
				 \justifying 
	\footnotesize{RDD local randomization: p-values are obtained from 1,000 permutations. Sharp RDD: conventional Eicker–Huber–White heteroskedasticity-robust standard errors. Source: Registrar General's Statistical Review.}
\end{figure}

Figure \ref{fig:2} presents compelling descriptive evidence of a significant decrease in IMR following the implementation of the NHS in 1948. It not only shows a sharp decline in IMR following the NHS introduction, but also aligns descriptively with major historical events, reinforcing the credibility of the data. For instance, in addition to a general decline in IMRs, they spiked in 1940–41, coinciding with the intense German bombing during WWII,\footnote{The bombing campaigns over British cities in 1940–41 strained healthcare infrastructure and living conditions, contributing to a 15.7\% increase in IMR compared to 1939 \citep{gardiner2010blitz}.} before dropping significantly in 1942, in line with German military focus being redirected toward the Soviet Union.\footnote{The so-called Operation Barbarossa, which began in June 1941 \citep{erickson2019road}.} Another decline occurred in 1946 following the war's end, and a final, notable drop appeared in 1948 at the point of NHS implementation. After this, the IMR trend became smooth and steadily declining—consistent with gradual improvements in living standards and public health.\footnote{Post-war improvements in sanitation and medical care contributed to the ongoing decline in IMR \citep{marshall2019mortality, mackenbach2020history}.} 

The table at the bottom of Figure \ref{fig:2} presents estimates and p-values from an RDD under local randomization, examining the causal effect of the NHS on infant mortality. Using a strict bandwidth of two years before and after the NHS implementation, the estimated effect is a reduction of 6.5 deaths per 1,000 births, corresponding to a 17.0\% decrease relative to the pre-NHS mean (38.44). Note that the two-year window excludes the WWII period, thereby avoiding potential contamination of the pre-NHS baseline.

Expanding the time window to six years before and after implementation yields a similar trend, with a larger estimated reduction of 10.17 deaths per 1,000 births, equivalent to a 24.16\% decrease relative to the pre-NHS mean (42.08). This estimate is also precisely estimated. However, the sharper decline observed with the wider bandwidth appears to be driven by contamination from the WWII years. For this reason, we prefer the specification that restricts the sample to two years on either side of the NHS introduction. The main results are consistent when estimating a sharp RDD model, as shown in the second column of the table under Figure \ref{fig:2}.

\subsection{Weekly Infant Mortality}
The previous section suggests that the introduction of the NHS substantially reduced IMR in England and Wales. However, that analysis is based on annual data, contaminating our treatment definition, and with that, reducing precision. The NHS was introduced in July 1948, and we consider the full year 1948 as the treatment period. As a result, our treatment group—intended to capture the discontinuity—is contaminated by births occurring in the six months preceding NHS implementation. Furthermore, our running variable is limited, consisting of only two observation points (the years before and after the NHS). The small number of observations constrains the precision of our estimates, while expanding the bandwidth beyond two year increases the risk of contamination from other contemporaneous events. 

\begin{figure}[H]
	\centering
	\includegraphics[width=1\textwidth]{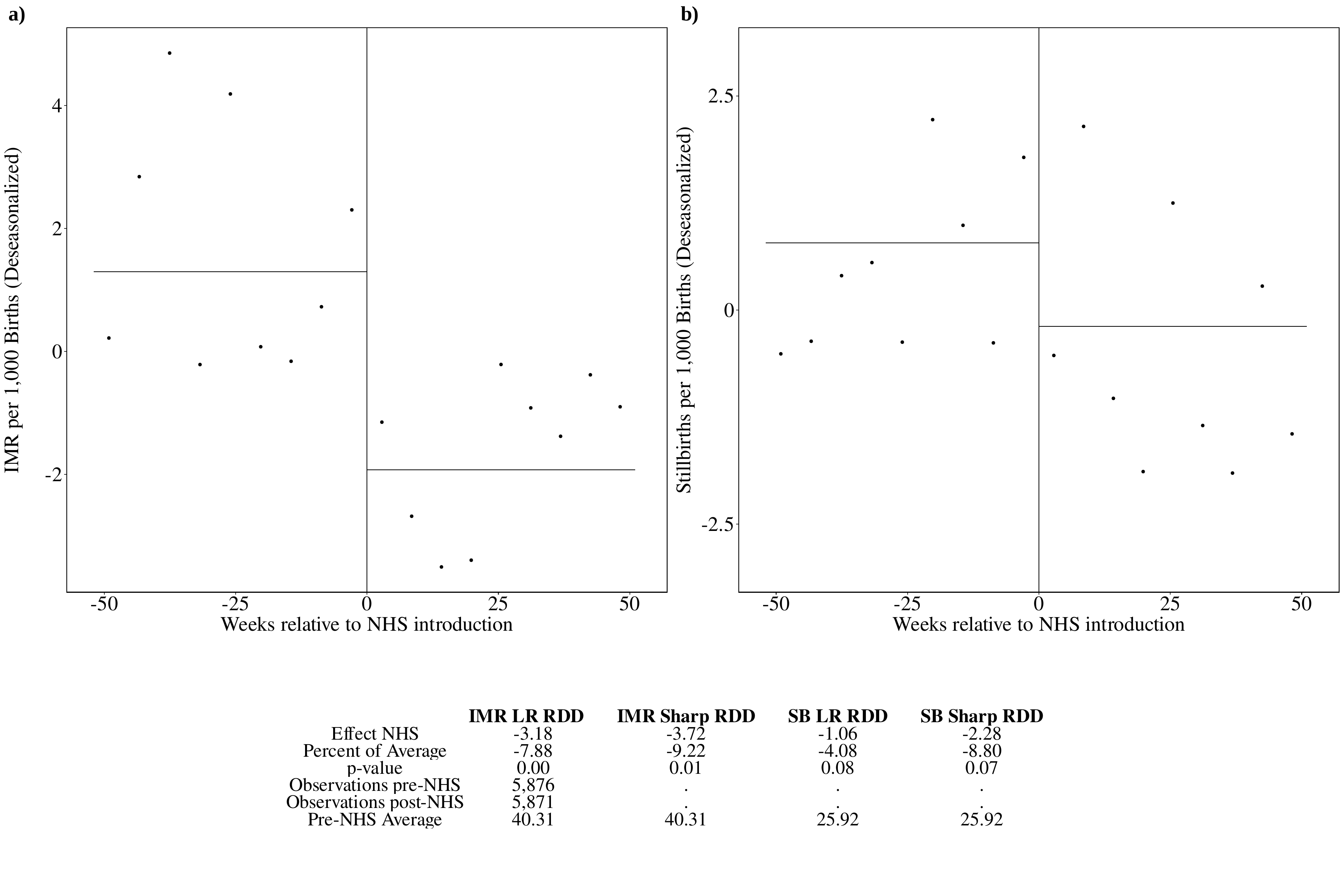}
	\caption{Effects of the NHS on Early-Life Mortality}
	\label{fig:weekly}
	\justifying 
	\footnotesize{RDD local randomization: p-values are obtained from 1,000 permutations. Sharp RDD: conventional Eicker–Huber–White heteroskedasticity-robust standard errors.
Source: Registrar General’s Weekly Returns. }
\end{figure}

To address these limitations, we explore weekly mortality data. Using a one-year bandwidth on either side of the threshold and collapsing the data to the monthly level to reduce measurement error, Figure \ref{fig:weekly} shows descriptive RDD graphs for two measures of mortality: the IMR in panel (a), and the rate of stillbirths (number of stillbirths per 1{,}000 live births) in panel (b). The data are residualized with respect to month and area fixed effects, and recentered on the date of the NHS introduction. 

Figure \ref{fig:weekly} presents a clear pattern for IMR: during the year before and after the NHS, the IMR fluctuates within a range of about five deaths per 1,000 live births. Crucially, following the NHS introduction, there appears to be a distinct downward shift. For stillbirths, the results are noisier but also suggest a discontinuity at the introduction of the NHS.

We estimate the RDD models under local randomization and confirm the descriptive patterns. The IMR decreases by 3.18 deaths per 1,000 live births in the local randomization RDD (an 7.88\% reduction relative to the pre-NHS mean) and by 3.72 deaths (9.22\%) in the sharp RDD. For stillbirths, the estimated effects are reductions of 1.06 deaths (4.08\%) and 2.28 deaths (8.80\%), respectively. The coefficients for IMR are statistically significant at the conventional 95\% confidence level, while those for stillbirths are significant at the 90\% level.

Although these results are smaller in magnitude compared to the yearly data since they do not cover the whole of England and Wales, they remain substantial. 
Overall, these results provide a precise picture of the NHS having substantially reduced stillbirths and IMR.

\subsection{Robustness Checks}


We perform a range of sensitivity analyses. First, we replicate the results using alternative bandwidths—three, six, 12, 18, and 24 months—as shown in Figure \ref{fig:local_rand_windows_IMR}. Across all choices, the estimates display the same overall pattern as in the baseline 12-month specification.

Second, in Figure \ref{fig:IMR_kernel}, we present results using different kernels. While the baseline specification uses a triangular kernel, we replicate the analysis using uniform and Epanechnikov kernels. The results are qualitatively and quantitatively similar across all kernels. We nevertheless favor the triangular kernel, as it is standard in the RDD literature and places more weight on observations closer to the cutoff, providing a more conservative and locally informative estimate \citep{cattaneo2022regression}. 

Finally, we estimate a doughnut specification in which observations closest to the NHS introduction date are excluded from the analysis. Specifically, we remove a three-month window on either side of the cutoff to avoid potential contamination from short-run irregularities or reporting noise around the implementation date. This approach ensures that the identifying variation comes from observations that are sufficiently removed from any immediate transition dynamics. In Figure \ref{fig:mortality_weekly_donut_IMR_SB_only} the results remain consistent with those obtained in the baseline specifications.

\subsection{Early-Life Mortality in Context}
We benchmark these magnitudes against prior work studying comparable health policy interventions. Exploiting the large Medicaid eligibility expansions in the US during the 1980s, \citet{currie1996saving} show that a 20–percentage-point increase in eligibility among women of childbearing age reduced infant mortality by around 7\% and significantly lowered the incidence of low birthweight, with substantially larger effects arising from earlier, more targeted expansions with higher take-up. \citet{goodman2018public} similarly find large declines in infant and child mortality following Medicaid expansions, with mortality among nonwhite children falling by roughly 20\% and aggregate nonwhite child mortality declining by about 11\%.

Evidence from Scandinavia points to similarly large mortality reductions following universal early-life health interventions. As summarized by \citet{wust2022universal}, in Denmark, the staggered rollout of a nationwide nurse-led home visiting program between 1937 and 1949 increased infant survival by 0.5--0.8 percentage points, corresponding to approximately 5--8 lives saved per 1{,}000 births, or a 9--17\% reduction in infant mortality. In Sweden, expanded access to professional midwife-assisted maternity care reduced infant mortality by about 1.6 percentage points, implying a 24\% decline in deaths during the first year of life. In Norway, comparable early-life health interventions were associated with a reduction in infant mortality of roughly 0.8 percentage points, corresponding to an 18\% decline.

Finally, although focused on adult rather than early-life mortality, \citet{bauernschuster2020bismarck} show that the introduction of Bismarck’s Health Insurance reduced mortality among blue-collar workers by approximately 8.9\% between its introduction and the turn of the twentieth century, accounting for roughly one-third of the total mortality decline observed for this group over the period.

Our preferred estimates---a \(17\%\) (yearly data) and \(7.9\%\) (weekly data) reduction in infant mortality relative to the pre-reform mean---are therefore of comparable magnitude to the effects documented across the US and Scandinavia.
In sum, the NHS enabled a significant group of individuals to survive at birth, potentially leading to selective survival. 


\section{Selective Survival} \label{sec:surv}
In this section, we examine whether the introduction of the NHS caused differences in PGI values in a local neighborhood around the eligibility cutoff. Genetic variation is determined at conception and, as such, PGIs are exogenous to the implementation of the NHS. However, the NHS could have influenced survival rates. If a sufficiently large number of individuals with distinct genetic traits survived after the implementation of the NHS, this could have left a genetic footprint, reflected by changes in the PGI distributions for those born pre- and post-NHS. 

\begin{figure}[htbp]
	\centering
		\includegraphics[height=0.7\textheight]{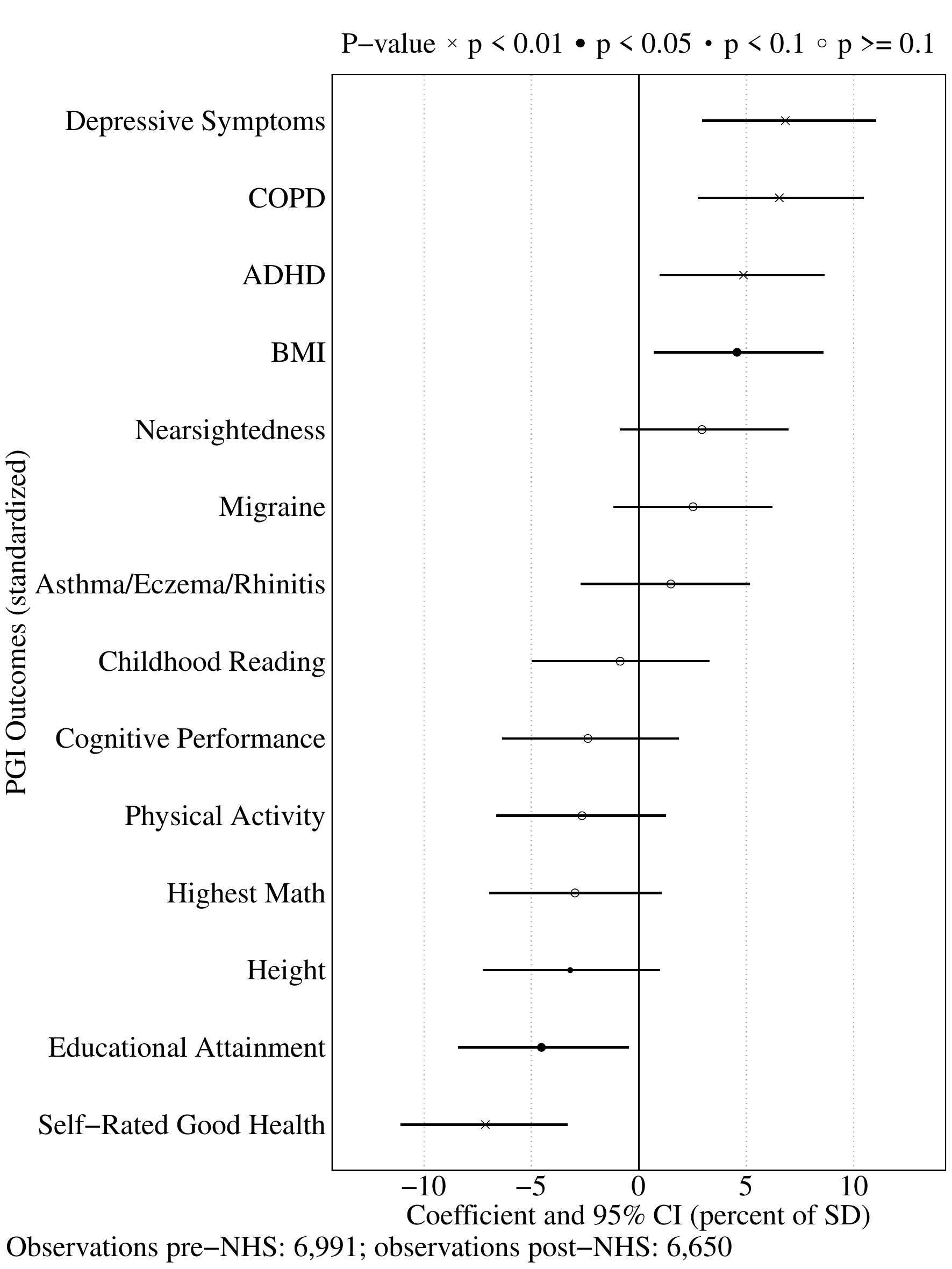}  
	\caption{Effects of the NHS on PGIs}
	\label{fig:3}
				 \justifying 
	\footnotesize{Note: Each dot represents the causal effect of a separate regression on a different outcome. P-values under permutation (1,000 permutations) and standard errors under interference. The bandwidth is 12 months pre- and post-NHS implementation. Source: UKB.}
\end{figure}

We begin with the results from our largest sample: the UKB. In Figure \ref{fig:3}, we present a forest plot illustrating the effects of the NHS on various PGIs. Each dot represents the coefficient from a separate regression, with a different PGI as the outcome, along with its p-values and 95\% confidence intervals. The results show a clear trend: the values of PGIs associated with contextually adverse conditions, such as depressive symptoms, COPD, or ADHD, increase after the implementation of the NHS, while the values for PGIs related to contextually favorable conditions, such as self-rated good health, educational attainment or height decrease. Effect sizes range from $\sim$-7.5\% to 7.5\% of a standard deviation. Even for PGIs in the middle range, although estimates are imprecise, the main coefficients point in a consistent direction, with contextually adverse conditions presenting positive effects and contextually favorable conditions negative ones.

\clearpage

These results suggest that a higher proportion of vulnerable individuals survived birth, thanks to the implementation of the NHS. These individuals were genetically different from the pre-NHS samples. We interpret this as evidence of the survival bias caused by the NHS.

\subsection{Robustness Checks} \label{sec:robustness}

\paragraph{Different specifications.} First, we replicate the results using different bandwidths, as shown in Figure \ref{fig:wind}. This includes bandwidths of three, six, 12, 18, and 24 months. In addition, Figure \ref{fig:fig2_bal} reports results from a covariate balance window selection procedure \citep{cattaneo2023practical}, which selects a bandwidth of seven months. Overall, the results maintain the same pattern as those estimated using the baseline 12-month bandwidth. The most noticeable trend is that, as the bandwidth narrows, the estimated effects become larger but also more imprecisely estimated due to the loss of statistical power. This is most evident with the three-month bandwidth, where, although the general pattern is preserved, effect sizes reach up to $\sim$10\% of a SD, and some estimates become statistically insignificant (ADHD, BMI, educational attainment). For this reason, we favor the more conservative specification that yields the most precisely estimated effects, namely, a bandwidth of 12 months. Furthermore, consistent with the analysis in Section \ref{fertility}, our results with different bandwidths (Figure~\ref{fig:wind}) suggest that the genetic selection effects cannot have been driven by fertility changes, since even the models with a short bandwidth of three or six months yield consistent results, albeit with larger standard errors. 

Second, in Figure \ref{fig:kern}, we present results using different kernels. While the baseline specification uses triangular kernels, we replicate the analysis using uniform and Epanechnikov ones. The results are qualitatively and quantitatively similar across all kernels.

Third, we replicate the main analysis using a sharp RDD with a linear specification. We fit separate linear trends on either side of the NHS implementation cutoff, allowing both the intercept and slope to differ across the threshold. The causal effect is estimated as the difference in the intercepts of these two regression lines at the cutoff point, capturing the discontinuity in the outcome variable attributable to the treatment. The model controls for month-year of birth as the running variable, interview date, quarter of birth fixed effects, gender, and ten ancestry-specific principal components. It is estimated with triangular kernel weights and standard errors using cluster-robust variance estimation with degrees-of-freedom adjustments, combined with nearest neighbor variance estimations. Figure \ref{fig:sharp} shows that the sharp RDD estimates closely mirror those obtained from the local randomization RDD. The primary distinction is that the sharp RDD estimates are notably larger, reaching up to 10\% of a SD. 

Fourth, we estimate a multivariate linear regression model where the treatment variable is defined as being born after the implementation of the NHS. The model includes covariates for date of birth, date of the interview, gender, quarter-of-birth fixed effects, and ten ancestry-specific PCs. Additionally, we perform two-sample t-tests to compare the mean differences in PGI values between individuals born before and after the NHS implementation. Figure~\ref{fig:pre_post} illustrates that both approaches yield results consistent with our baseline specifications, exhibiting similar qualitative and quantitative patterns. The similarity in effect sizes suggests that these mean comparison methods are inherently similar to the local randomization RDD approach.

\paragraph{Multiple hypothesis testing.} We address multiple hypothesis testing in Figure~\ref{fig:mult} by applying both the Benjamini-Hochberg procedure \citep{benjamini1995controlling} and the Bonferroni correction \citep{dunn1961multiple}. The Benjamini-Hochberg method controls the false discovery rate, offering a balance between identifying true effects and limiting false positives. In contrast, the Bonferroni correction controls the family-wise error rate, providing a more conservative adjustment that reduces the likelihood of any false positives but may increase the chance of overlooking true effects. Applying the Benjamini-Hochberg adjustment, only one outcomes lose statistical significance compared to the baseline specification: the PGIs for height. When using the more conservative Bonferroni correction, several additional outcomes lose significance. However, self-rated good health, depression, and COPD remain statistically significant at the 99\% confidence level. Regardless of statistical significance thresholds, it is important to note the consistent qualitative gradient observed across phenotypes. This pattern suggests a directional selection, reinforcing the validity of our findings beyond mere statistical precision.

\paragraph{Full set of PGIs.} We extend our analysis to include all available PGIs from the Polygenic Index Repository. These encompass PGIs derived from both single- and multiple-imputation GWAS, as well as phenotypes related to fertility, sexual development, health behaviors, and personality traits. Figure \ref{fig:allPGI} illustrates that the results obtained using single-input and multiple-input GWAS PGIs are nearly identical, indicating that the choice of GWAS method does not significantly affect the outcomes. Furthermore, the analysis reveals that only a few of the additional outcomes exhibit statistically significant effects. Notably, these significant associations pertain to personality traits, such as loneliness and neuroticism, which exhibit higher values following the NHS implementation. This aligns with our previous findings, considering these phenotypes as contextually detrimental and associated with poor mental health.

\paragraph{Different samples.}
We replicate our baseline analysis using data from ELSA and USoc, which, although smaller in size, are nationally representative. The set of PGIs in ELSA differs from those in the UKB and is based on \citet{ajnakina2022english}. Nevertheless, we select PGIs equivalent to those used in the UKB, namely in the categories of socioeconomic traits, adult mental health and well-being, physical health, anthropometric traits, and intelligence. For USoc, we construct the same PGIs as in the UKB analysis based on \citet{becker2021resource}.

Figure \ref{fig:elsa} shows that although the results with the ELSA sample are much more imprecisely estimated, the previously described qualitative patterns remain. PGIs related to contextually detrimental conditions such as migraine and schizophrenia show higher values after the implementation of the NHS, while phenotypes such as IQ, general cognition, and educational attainment (EA3) show lower ones. Estimates are noisy, with coefficient sizes ranging from -25\% to 25\% of an SD, largely due to the smaller sample size of only 705 observations. Therefore, we believe that the true effect sizes are closer to those in our baseline specification ($\pm5\%$ of a SD).

Finally, the results for USoc are presented in Figure \ref{fig:USoc}, showing estimates that are even more imprecisely measured. Nevertheless, the qualitative pattern remains consistent, and the only PGIs with  negative significant effects are height and self-reported good health, the latter of which similarly shows the largest negative effect in the UKB.

\section{Understanding selective survival} \label{sec:selective_survival}

Our findings reveal a drop in early-life mortality which is indeed selective. In this section, we aim to understand the mechanisms driving these  effects. Firstly, we  focus on different causes of death to explain how survival is enhanced. Secondly, we unpack the selective effects by studying (i) whether they operate through within- or between-family pathways employing family fixed-effects designs, (ii) whether they are captured by different types of PGIs related to early-life conditions, and (iii) whether effects vary by gender testing the the male frailty hypothesis. 
Finally, we discuss potential key changes in healthcare usage and supply induced by the NHS, and we show heterogeneities by SES.

\subsection{Causes of Death}

\begin{figure}[H]
	\centering
	\includegraphics[width=1\textwidth]{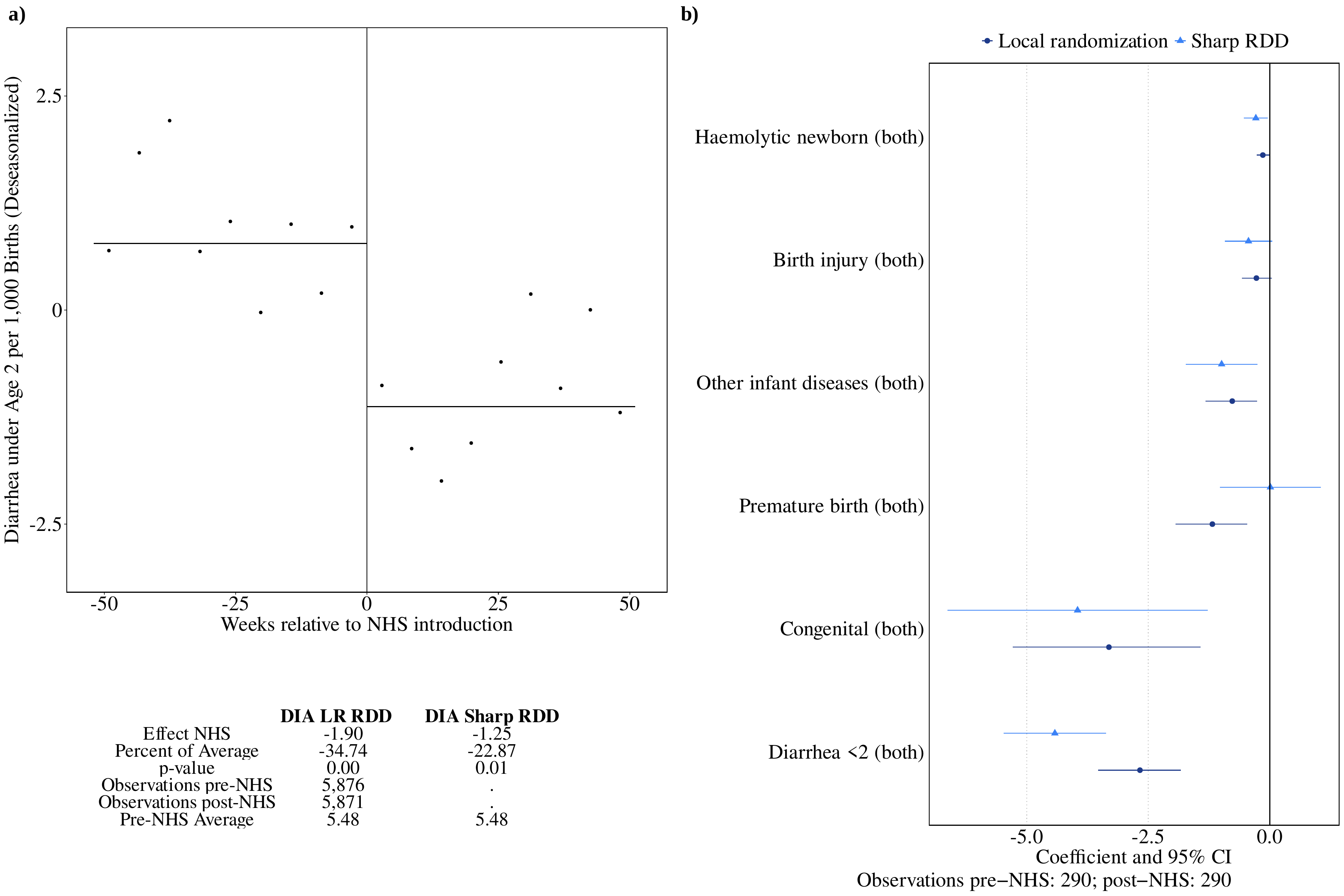}
	\caption{Effects of the NHS on Early-Life Mortality by Causes}
	\label{fig:weekly_cause}
	\justifying 
	\footnotesize{Note: Each dot represents the causal effect from a separate regression on a different outcome. RDD local randomization: p-values are obtained from 1,000 permutations, and confidence intervals use interference-robust standard errors. Sharp RDD: conventional Eicker–Huber–White heteroskedasticity-robust standard errors. The bandwidth is 52 weeks in panel (a) and two years in panel (b) before and after the NHS implementation. Source: Registrar General’s Weekly Returns for panel (a) and Registrar General’s Statistical Reviews for panel (b).}
\end{figure}

To understand how the introduction of the NHS affected mortality, we first examine which causes of death contributed to the reductions in IMR and stillbirths. Using the \textit{Registrar General’s Weekly Returns}, which report statistics for urban areas in England and Wales, we focus on deaths from diarrhea among children under age two. Figure~\ref{fig:weekly_cause}, panel~(a), shows that the NHS reduced  diarrhea mortality by 1.90 deaths per 1{,}000 live births (34.74\%). Although this measure includes deaths up to age two (rather than only those in the first year of life), this effect corresponds to roughly 40.25\% of the weekly IMR reduction shown in Figure~\ref{fig:weekly}.\footnote{The corresponding sharp RDD estimate implies a reduction of 1.25 deaths per 1{,}000 live births (22.87\%).}

Second, we use the \textit{Registrar General’s Statistical Reviews}, which report annual mortality by cause, with some causes being specific to the infancy period. In addition to mortality from diarrhea under age two, we examine effects on congenital conditions, premature births, birth injuries, haemolytic disorders of the newborn, and other infant diseases. Figure~\ref{fig:weekly_cause}, panel~(b), shows that all coefficients are negative, indicating consistent reductions in infant mortality across causes. Under the local randomization specification, the largest reductions are observed for diarrhea under age two and for congenital conditions: diarrhea mortality fell by 2.67 deaths per 1{,}000 live births (a 53\% decline relative to the pre-NHS mean), and congenital mortality fell by 3.31 deaths per 1{,}000 (a 9.7\% decline). Premature birth mortality declined by 1.18 deaths per 1{,}000 live births (10.8\%), and mortality from other infant diseases declined by 0.77 deaths per 1{,}000 (16.5\%). Birth injuries and haemolytic disorders show smaller effects—0.28 deaths per 1{,}000 (10\%) and 0.14 deaths per 1{,}000 (19.9\%), respectively—reflecting the low baseline mortality rates for these conditions. Most estimates are statistically significant at the 95\% confidence level.\footnote{Sharp RDD estimates are generally larger in magnitude across causes, with premature birth being the only exception.} Overall, the reductions typically range from 10–20\% of the baseline mortality rate, with diarrhea again standing out as the condition with the largest proportional decline.

Finally, we repeat these analyses for maternal deaths at delivery. These outcomes include deaths due to childbirth-related infections, overall maternal mortality, maternal sepsis, abortion-related causes (septic and non-septic), and a separate category capturing other specified maternal complications such as hemorrhage or embolism. Figure~\ref{fig:maternal} in the Appendix presents suggestive evidence of moderate reductions in overall maternal mortality and in the other maternal complications category. Nevertheless, these estimates are statistically significant only under the local randomization approach, with smaller coefficients under the sharp RDD. For childbirth-related infections, septic abortion, and maternal sepsis, the estimated coefficients are close to zero.

\subsection{Within-Family Results}

We next explore whether our analysis is robust to holding the family environment constant and exploiting variation in NHS exposure and PGIs only within families. We do this in two ways, estimating specifications that control for siblings’ PGIs--which proxy for familial genetic nurture--and, more restrictively, include family fixed effects--to absorb shared family-level genetic background and household environments. 

\begin{figure}[H]
	\centering
	\includegraphics[width=15cm]{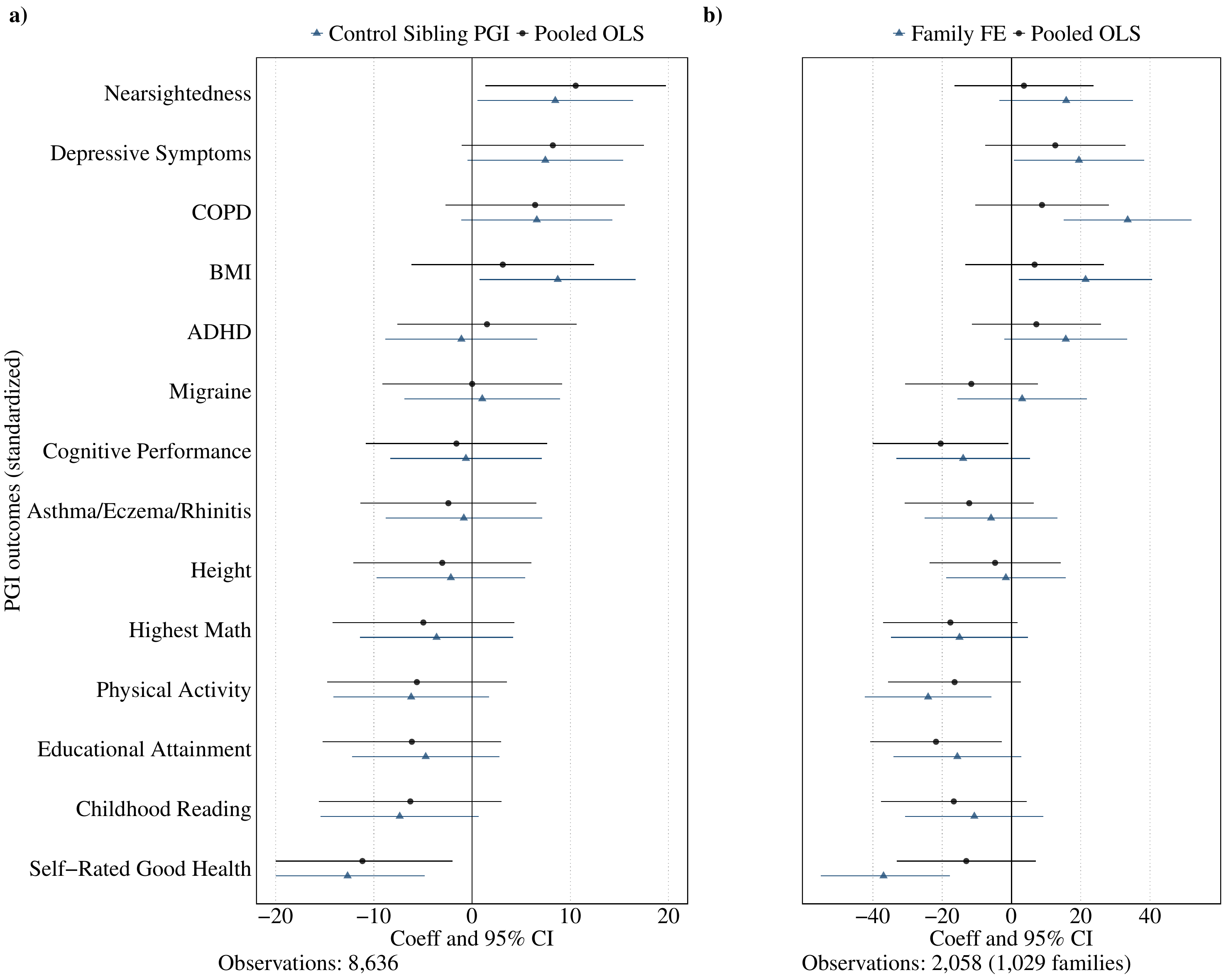}
	\caption{Within-Family Effects of the NHS on PGIs}
	\justifying 
	\label{fig:siblings}
	\footnotesize{Note: Each dot represents the causal effect of a separate regression on a different outcome. Clustered standard errors at the family level. The bandwidth is 12 months pre- and post-NHS implementation. Controls include continuous year-month of birth, interview date (age), gender, quarter of birth fixed effects, and the first 10 ancestry-specific PCs. We additionally control for the gender of each respective sibling.  Source: UKB.}
\end{figure}

Figure \ref{fig:siblings} presents these estimates, with panel (a) showing results from pooled models controlling for siblings’ PGIs and panel (b) reporting specifications that include family fixed effects. Although the coefficients are less precise and most confidence intervals overlap zero, the previously observed pattern remains clearly visible across specifications. Interestingly, the pooled models and those controlling for siblings’ PGIs or family fixed effects yield effect sizes of similar magnitude. In some cases—such as the BMI PGI in the specification controlling for siblings’ PGIs, or COPD, BMI, and self-rated good health in the specification with family fixed effects—the estimated effect sizes even increase once family-level controls are included. 


These estimates are large in magnitude, in some cases reaching up to 30\% of an SD, but they are substantially less precise due to the smaller sibling sample. 
The specifications controlling for siblings’ PGIs include 8,636 individuals, while the family fixed-effect models rely on 2,058 sibling pairs (1,029 families). As a result, confidence intervals are wide and point estimates should be interpreted with caution.

Nevertheless, the qualitative pattern is remarkably stable. Even when comparing one sibling to another in the same family, the one who is exposed to the NHS around birth is more likely to have a lower PGI for contextually-valued traits and vice versa for contextually-adverse traits.

This persistence of the pattern within families suggests that the observed genetic selection is not driven solely by differences across families or by compositional changes at the household level. Instead, the results are consistent with selection operating at the individual level, potentially through biological pathways that affect survival in utero or in early-life and that differ even among siblings exposed to similar family environments.

\subsection{Early-life PGIs}

A key assumption of our analysis is that survival at birth, closely tied to early-life vulnerability, can be indirectly captured by examining phenotypes that manifest later in life. This approach relies on the assumption that the SNPs included in the PGIs under consideration are relevant to in-utero resilience. However, a more direct strategy is to focus on phenotypes that are explicitly linked to early-life survival. To this end, we construct PGIs using summary statistics from the Early Growth Genetics (EGG) Consortium.

We construct PGIs for birth weight \citep{beck2021genetic,warrington2019maternal}, head circumference \citep{vogelezang2022genetics}, and different measures of gestational duration \citep{sole2023genetic}. For each phenotype, we rely on the most recent available GWAS, with the exception of birth weight, where the latest study \citep{beck2021genetic} is based on a twin design. Twins differ systematically from the general population, particularly in traits such as birth weight, which is highly correlated within twin pairs. Therefore, we also use birth weight summary statistics from \citet{warrington2019maternal}.\footnote{
Since that GWAS includes the UKB in its discovery sample, we contacted the authors and obtained an alternative version of the summary statistics that excludes the UKB. } 

\begin{figure}[H]
	\centering
\includegraphics[width=11cm]{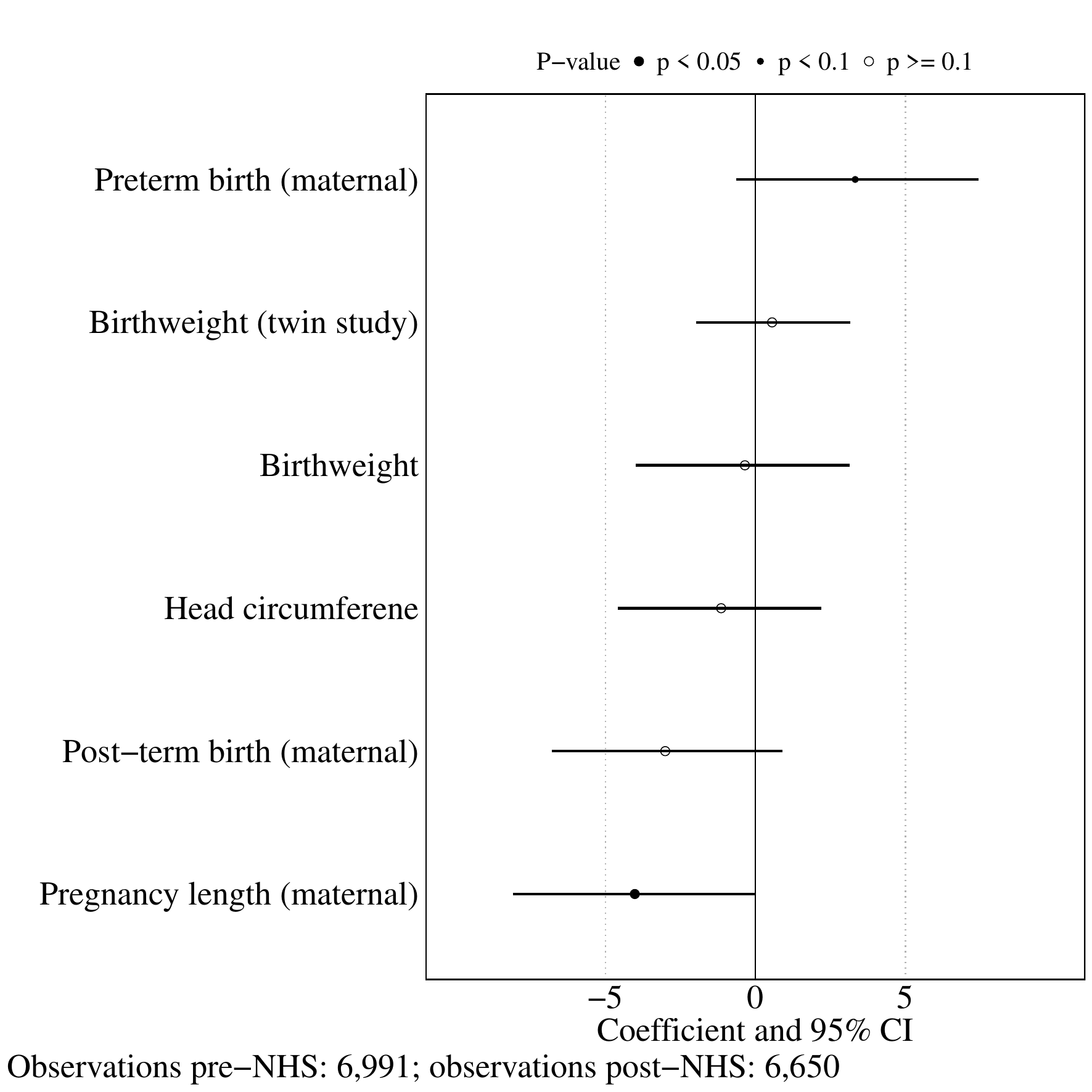}
	\caption{ Effects of the NHS on `early-life' PGIs}
	\label{fig:other_PGIs}
	\justifying 
	\footnotesize{Note: Each dot represents the causal effect of a separate regression on a different outcome. P-values under permutation (1,000 permutations) and standard errors under interference. The bandwidth is 12 months pre- and post-NHS implementation. Source: UKB.}
\end{figure}

Figure~\ref{fig:other_PGIs} reports the estimated effects of the NHS reform on PGIs capturing early-life and perinatal conditions. The clearest and most precisely estimated effects relate to maternal gestational outcomes. We find an effect of 3.33\% of a SD on the maternal preterm-birth PGI and -4.02\% on the maternal pregnancy-length PGI, significant at the 10\% and 5\% levels, respectively. Longer gestation are generally interpreted as markers of better early-life health, whereas shorter gestation and preterm birth are associated with worse neonatal and long-run outcomes \citep{currie2011human}. Taken together, these estimates are consistent with a shift toward contextually detrimental phenotypes and away from contextually beneficial ones following the implementation of the NHS.

Overall, the estimates are noisier, which is consistent with lower statistical power, as the underlying GWASs for these traits rely on substantially smaller discovery samples than the GWASs used for our baseline PGIs. Therefore, our preferred PGIs remain those based on \cite{becker2021resource}.

\subsection{The Male Frailty Hypothesis}
The male frailty hypothesis posits that boys follow a faster growth strategy in utero, which makes them more vulnerable to maternal stress. Male fetuses grow more rapidly and place greater physiological demands on the mother and placenta, yet their placentas are less adaptable to adverse environments and have less reserve capacity. When maternal stress occurs, boys are therefore more likely than girls to be miscarried, stillborn, or to experience long-term developmental disadvantages. Girls, by contrast, grow more slowly and exhibit greater placental plasticity, allowing them to buffer the effects of stress more effectively. The ultimate manifestation of these differences is that stressful environments often reduce the proportion of male births, consistent with the notion of male biological vulnerability in early-life \citep{kraemer2000fragile,eriksson2010boys}.

\begin{figure}[htbp]
	\centering
\includegraphics[width=12cm]{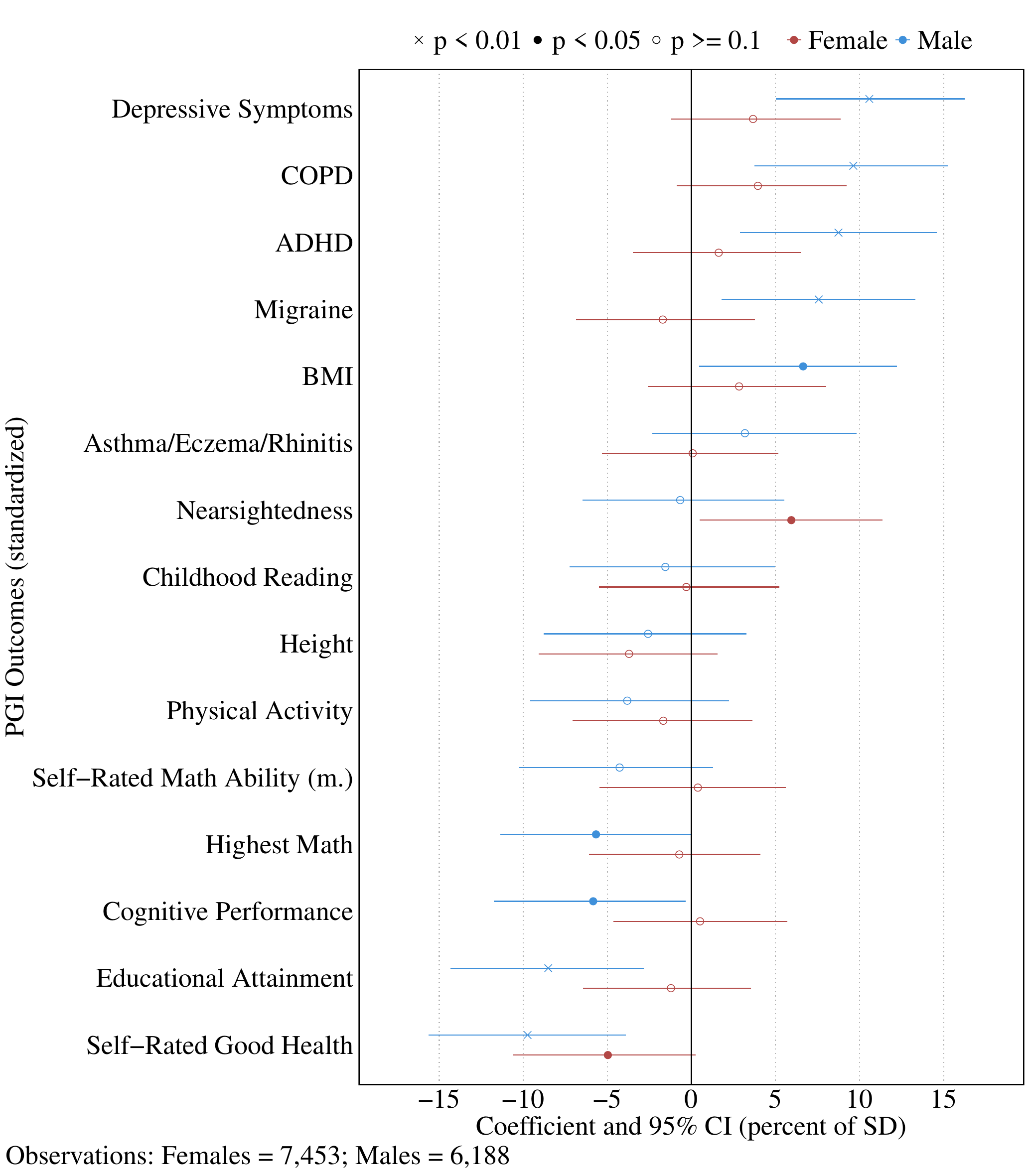}
	\caption{Heterogeneity by Gender}
	\label{fig:gender}
	\justifying 
	\footnotesize{Note: Each dot represents the causal effect of a separate regression on a different outcome. P-values under permutation (1,000 permutations) and standard errors under interference. The bandwidth is 12 months pre- and post-NHS implementation. Source: UKB.}
\end{figure}

According to this, and if the most vulnerable individuals were those who benefited the most from the introduction of the NHS, we would expect the effect to be larger for males than for females. In Figure \ref{fig:gender}, we stratify the analysis by gender. The results are suggestive of larger effects for males: in 12 out of 14 PGIs, the estimated effects are substantially larger for males, with one trait showing negligible differences and another (nearsightedness) showing larger effects for females. The latter may reflect biological differences specific to that PGI, but could also arise by chance, since we are estimating 28 separate hypotheses (14 PGIs stratified by gender), which increases the likelihood of spurious differences. Yet, for the traits with the largest and smallest estimated effects, the magnitudes for males are approximately 50\% greater than those for females. Although these gender differences are not statistically significant, the consistent pattern suggests that male fetuses were more affected by the introduction of the NHS in early-life than female fetuses, consistent with the male frailty hypothesis.

\clearpage

\subsection{Mechanisms}
An important question that follows from our analyses relates to the \textit{mechanisms} driving the effects on infant mortality and the genetic composition of the population. As highlighted in Section \ref{sec:IMRmechanism}, any reductions in infant mortality are driven by one of two channels that shape the demand for healthcare: a \textit{health} channel (e.g., access to skilled medical professionals, sterile environments) or an \textit{income} channel (e.g., reduced out-of-pocket spending). Although we cannot empirically distinguish between these mechanisms, both relate to a patient-response. The provider-response however, or the healthcare supply channel, is likely to be equally important. Although the introduction of the NHS did not coincide with an increase in the number of doctors or nurses, there may have been prioritization of those in need. Indeed, if medical practitioners gave preferred access to certain patients, the benefits of the improved health and reduced financial burden would disproportionally lie with those mothers and children.

To inform the discussion, we draw on historical public health documents published around the time of the NHS introduction. 
We use these to descriptively understand how maternity services were allocated \textit{within} a local area. 
A key factor discussed in Department of Health reports is on how to prioritize maternity services in case the demand for care exceeds supply, which was the situation in most areas. Such prioritization was done based on two grounds, as highlighted in a 1951 memorandum sent by the Ministry of Health to hospital authorities. This memorandum reflected the advice the Minister received from the Standing Maternity and Midwifery Advisory Committee on the selection of patients for hospital confinement \citep[Appendix IV,][]{cranbrook1959report}. The first prioritization was done according to medical or obstetric need. This was interpreted in the widest sense of these terms, including first (but also fourth and higher) births, multiple pregnancies, and patients with a history of stillbirths and neonatal deaths. The second prioritization included mothers living in adverse social conditions, especially those in bad housing. The report acknowledges that it is not possible to define criteria for ``adverse social conditions'', so in practice, these were assessed by local health visitors, district nurses or midwives who report back to the hospital on the suitability of mothers' dwellings \citep[see e.g.,][]{MoH1949Bristol,MoH1949Bradford,MoH1949Buckinghamshire,MoH1952Devon}. 

In short, this suggests that the introduction of the NHS is likely to have improved child health and reduced out-of-pocket spending disproportionally among poorer households, with hospitals giving priority to those in most need. Our final analysis explores whether the empirical evidence is consistent with this, investigating whether NHS-driven selection is stronger in lower socioeconomic status (SES) areas.

\subsection{Socioeconomic Heterogeneity}
\begin{figure}[H]
	\centering
\includegraphics[width=12cm]{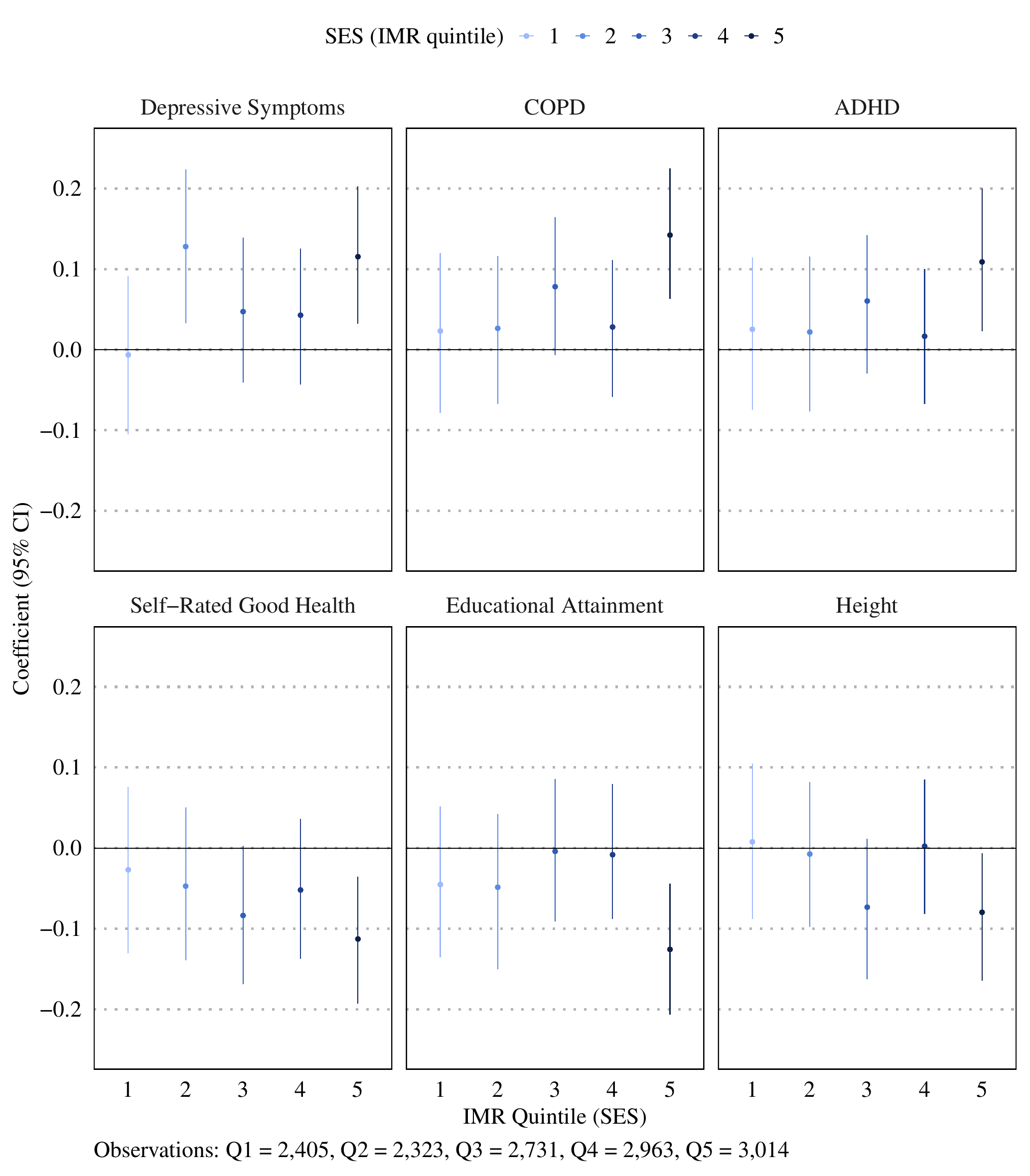}
	\caption{Heterogeneity by pre-NHS IMR}
	\label{fig:heterogeneity}
	\justifying 
	\footnotesize{Note: Each dot represents the causal effect of a separate regression on a different outcome. Standard errors under interference. The bandwidth is 12 months pre- and post-NHS implementation. Source: UKB.}
\end{figure}

In the absence of individual-level SES data around birth, we use a number of socio-economic characteristics measured at the level of individuals' district of birth around the implementation of the NHS. We start by using district-level infant mortality rates from 1947.
In Figure \ref{fig:heterogeneity}, we study if the impact of the NHS introduction on PGIs varies by quintiles of the IMR distribution where quintile one (five) corresponds to the areas with the lowest (highest) IMR. We focus on the top three contextually-beneficial and bottom three contextually-detrimental phenotypes. The results indicate a clear heterogeneity: the higher the IMR quintile the lower the value of the PGI for self-rated good health, height and educational attainment, and the higher the value of the PGIs for depressive symptoms, COPD and ADHD. This difference is substantial looking at the 5th IMR quintile, with effect sizes nearly doubling, suggesting that these more localized effects are driving the overall average treatment effects (ATEs).
When considering a larger set of PGIs, we draw the same conclusions (Appendix Figure \ref{fig:heterogeneity_all}). As a further analysis, we use the district-level social class information from the 1951 Census and perform a similar heterogeneity exercise.\footnote{The 1951 Census provides occupation-based social class measures following the Registrar General’s classification: Class I (professional), Class II (managerial and technical), Class III (skilled, manual and non-manual), Class IV (partly skilled), and Class V (unskilled). We use these categories to construct quintiles of district-level SES, defining quintile one (five) as districts with the lowest (highest) proportion of individuals in Classes IV--V.} The results are reported in Appendix Figure \ref{fig:heterogeneity_census}, 
confirming that the changes in IMR—and with that, mortality selection—are driven by low-SES districts (i.e., the top quintiles of the distribution). In other words, the introduction of the NHS impacted outcomes in predominantly low-SES areas, leaving higher-SES areas generally unaffected. This is consistent with Department of Health reports published around the time of NHS implementation, highlighting the prioritization of mothers who live in adverse social conditions; those more common in lower social class areas.

\section{Discussion}\label{sec:conclusion}
This paper shows new evidence on the impact of the introduction of the UK National Health Service (NHS) in 1948. Specifically, we use data from four sources 
to show that the introduction of the NHS significantly altered infant survival and, consequentially, the genetic composition of the surviving population. Our regression discontinuity design under local randomization shows that the NHS implementation caused a substantial decline in infant mortality rates (IMR), driving systematic shifts in polygenic indexes associated with both adverse and beneficial traits. 
NHS exposure increased PGIs for contextually adverse conditions (e.g., depression, chronic obstructive pulmonary disease) and decreased PGIs for contextually advantageous traits (e.g., educational attainment, self-rated good health).

Our findings make three core contributions. First, we contribute to the literature investigating the impact of national health policies. The NHS is the first fully universal public health system of the 20th century \citep{rivett1998cradle} and we show that it significantly reduced infant mortality. We analyze potential transmission channels of this effect. Historical reports suggest that these effects are at least partially driven by a prioritization in the NHS of more vulnerable pregnant women, not only on medical or obstetric grounds, but also along social and environmental dimensions, such as poor housing and unsanitary conditions. Our evidence is consistent with this, showing improved survival among higher-risk individuals and increased protection for individuals born in disadvantaged areas.

Second, we contribute to the existing literature on early-life conditions \citep{almond2018childhood, goodman2021long} by identifying and quantifying the selective survival associated with a major universal health policy. Using genetic data and a robust causal econometric methodology, we investigate changes in the genetic profile of the population before and after the introduction of the NHS, and show that NHS exposure increased the prevalence of genetic endowments contextually associated with adverse conditions, such as COPD and depression, and decreased those linked to educational attainment and self-rated good health. The magnitudes of these effects are consistently around 7.5\% of a standard deviation, and are robust across a wide range of specifications.

Third, our analysis contributes to the rapidly evolving literature on social-science genomics. While earlier studies have relied on associational designs or focused on single phenotypes \citep{zhou2024genetic, furuya2024separating}, we combine causal analysis of a policy change with a broad set of PGIs to trace the genetic footprint of a historical intervention. 

Our findings have substantive implications for both population genomics and human genetic evolution. Groundbreaking work in ancient and ancestral genomics has revealed that genetic change throughout human history and prehistory occurred in a more episodic and dynamic manner than previously thought \citep{reich2018we, skoglund2016genomic}. These shifts were driven by the strength and direction of natural selection—one of the central mechanisms of genetic evolution \citep{darwin1950origin,fisher1958genetical}—which, in turn, was shaped by major environmental and societal disruptions. Episodes of intense selection, often linked to large-scale migrations, exposure to infectious diseases, or transitions in diet and subsistence, left measurable and lasting signatures in the human genome. Yet even such substantial evolutionary shifts unfolded over centuries or millennia, underscoring the relatively slow pace of genomic adaptation compared with cultural or environmental change \citep{lazaridis2014ancient, mathieson2015genome, reich2018we}. 

In contrast to this long history of gradual genetic change, the past two centuries have seen an extraordinary acceleration in human development and environmental transformation. Since the Industrial Revolution, global GDP per capita has increased more than tenfold, life expectancy has more than doubled, and levels of education, sanitation, and healthcare have improved at an unprecedented pace \citep{comin2010exploration, comin2018if, deaton2006global}. These advances have reshaped nearly every aspect of human life—altering survival, fertility, and the distribution of environmental exposures within and across populations—and it is therefore to be expected that they left a lasting footprint.
Our analysis of one of the most influential healthcare policies of the twentieth century provides empirical evidence consistent with this view, suggesting that genetic shifts can indeed be detected at the population level and may occur in remarkably sharp and pronounced ways. This is particularly relevant given that contemporary societal transformations—such as climate change, air pollution, and the rise of digital and AI-driven environments—are likely to impose even swifter and more intense selective pressures on human populations.

Our findings also have important implications for empirical research on early-life interventions, cautioning against interpreting estimated treatment effects as fully representative of the treated population. Observed outcomes may reflect both a \emph{long-term scarring effect}—the direct impact of an early-life shock on individual outcomes—and a \emph{long-term selection effect}—changes in who remains in the sample due to selection \citep{nobles2019detecting}. In most settings, the primary parameter of interest is the long-term scarring effect \citep{currie2015early}. However, because the observed sample is subject to non-random selection, empirical estimates may conflate scarring and selection. These two potentially opposing forces can interact in complex ways, complicating causal interpretation. In our context, the NHS altered the genetic composition of the baseline population, implying that subsequent estimates of its long-term effects on health, education, or labor market outcomes will partly capture selection.  

Crucially, our results provide an empirical measure of the magnitude of genetic selection in terms of deviations in the genetic structure of the population. This should make it possible to empirically approximate the extent to which selection mechanisms may bias causal estimates of long-run effects. By leveraging the predictive capacity of genetic signals for phenotypic traits and combining this information with selection effects, differences in long-term outcomes could be approximated as arising from underlying variation in genetic composition. This opens the door to methodological innovation aimed at correcting such bias by reweighting unbalanced samples using genomic data through approaches such as entropy balancing or propensity score matching \citep{hainmueller2012entropy, caliendo2008some}, or by formally bounding the potential bias using partial-identification methods such as Lee bounds \citep{lee2009training}. Together, these approaches highlight a path forward for integrating genetic data into causal inference frameworks—allowing for more accurate estimation of both direct and selection-driven components of long-term policy effects.

In sum, our study provides direct evidence of the immediate impact of universal public health coverage in terms of infant mortality, and it shows that this type of policy shock can leave a genetic imprint on the surviving population. We contribute to the historical evaluation of the NHS and demonstrate that selective survival can be empirically identified and quantified — an insight that has broad implications for population genomics and is essential for credible and accurate assessments of early-life policies and shocks.


\newpage		
\bibliographystyle{apacite}
\bibliography{References}


\begin{appendices}	
	\onehalfspacing

	\onehalfspacing
\renewcommand{\thetable}{A.\arabic{table}}
\renewcommand{\thefigure}{A.\arabic{figure}}
\renewcommand{\thesection}{A}
\setcounter{figure}{0}
\setcounter{table}{0}
\section{Extra Results: Survival Bias}

\begin{figure}[H]
	\centering
	\includegraphics[height=0.5\textheight]{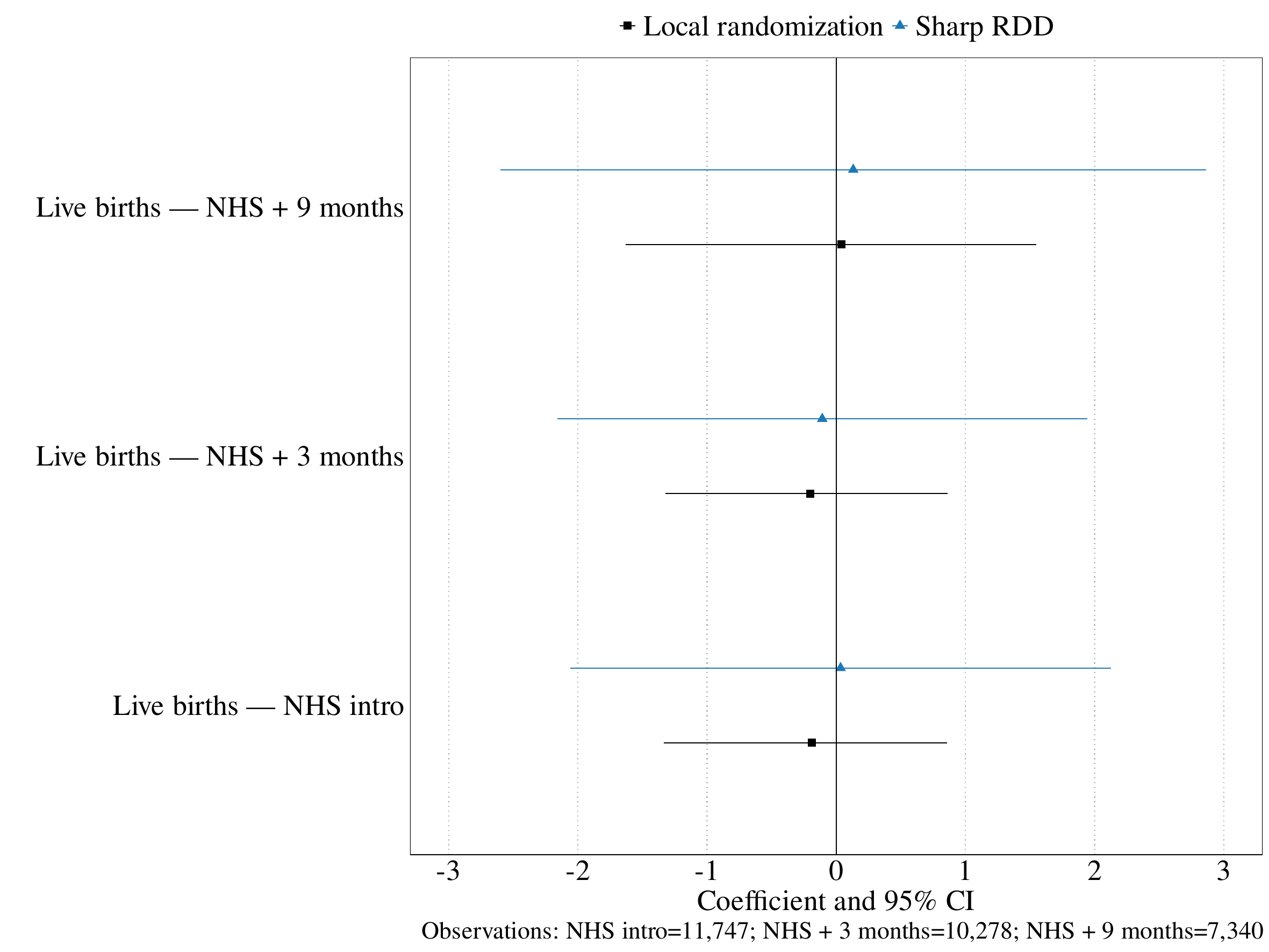}  
	\caption{ Effects of the NHS on Fertility}
	\label{fig:fertility1}
	\justifying 
	\footnotesize{Note: Each dot represents the causal effect of a separate regression on a different outcome. RDD local randomization: confidence intervals use interference-robust standard errors. Sharp RDD: conventional Eicker–Huber–White heteroskedasticity-robust standard errors. The bandwidth is defined as 12 months prior to the introduction of the NHS, while for the post-reform period we use alternative bandwidths: (i) 12 months after implementation, (ii) 4–15 months after implementation, and (iii) 10–18 months after implementation. Source: Registrar General’s Weekly Returns and Registar General of Statistics.  }
\end{figure}

\begin{figure}[H]
	\centering
	\includegraphics[height=0.7\textheight]{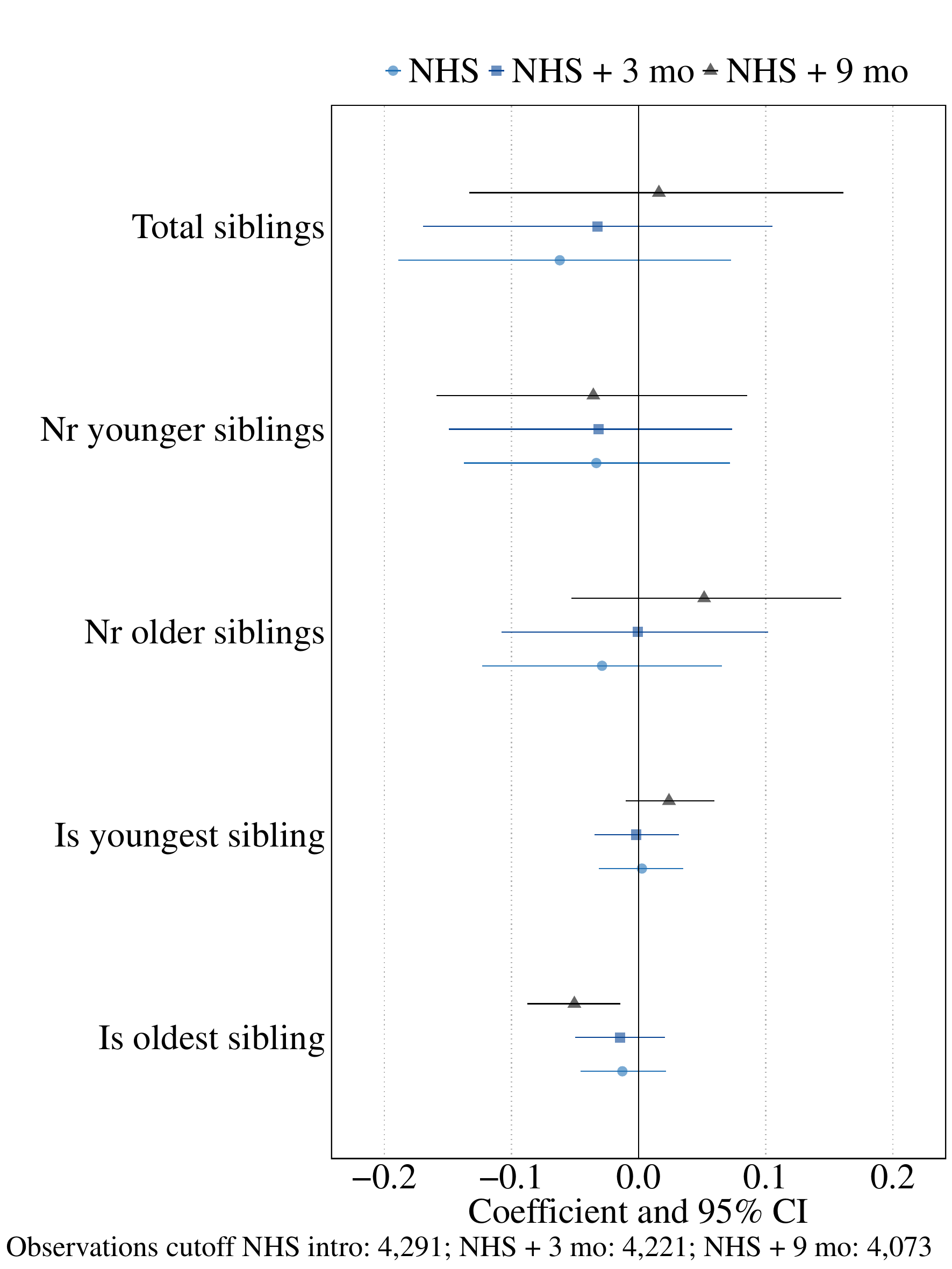}  
	\caption{ Effects of the NHS on Family Structure}
	\label{fig:fertility_smallsample}
	\justifying 
	\footnotesize{Note: Each dot represents the causal effect of a separate regression on a different outcome.RDD local randomization: confidence intervals use interference-robust standard errors. Sharp RDD: conventional Eicker–Huber–White heteroskedasticity-robust standard errors.. The bandwidth is defined as 12 months prior to the introduction of the NHS, while for the post-reform period we use alternative bandwidths: (i) 12 months after implementation, (ii) 4–15 months after implementation, and (iii) 10–18 months after implementation. Source: UKB.}
\end{figure}

\begin{figure}[H]
	\centering
	\includegraphics[height=0.7\textheight]{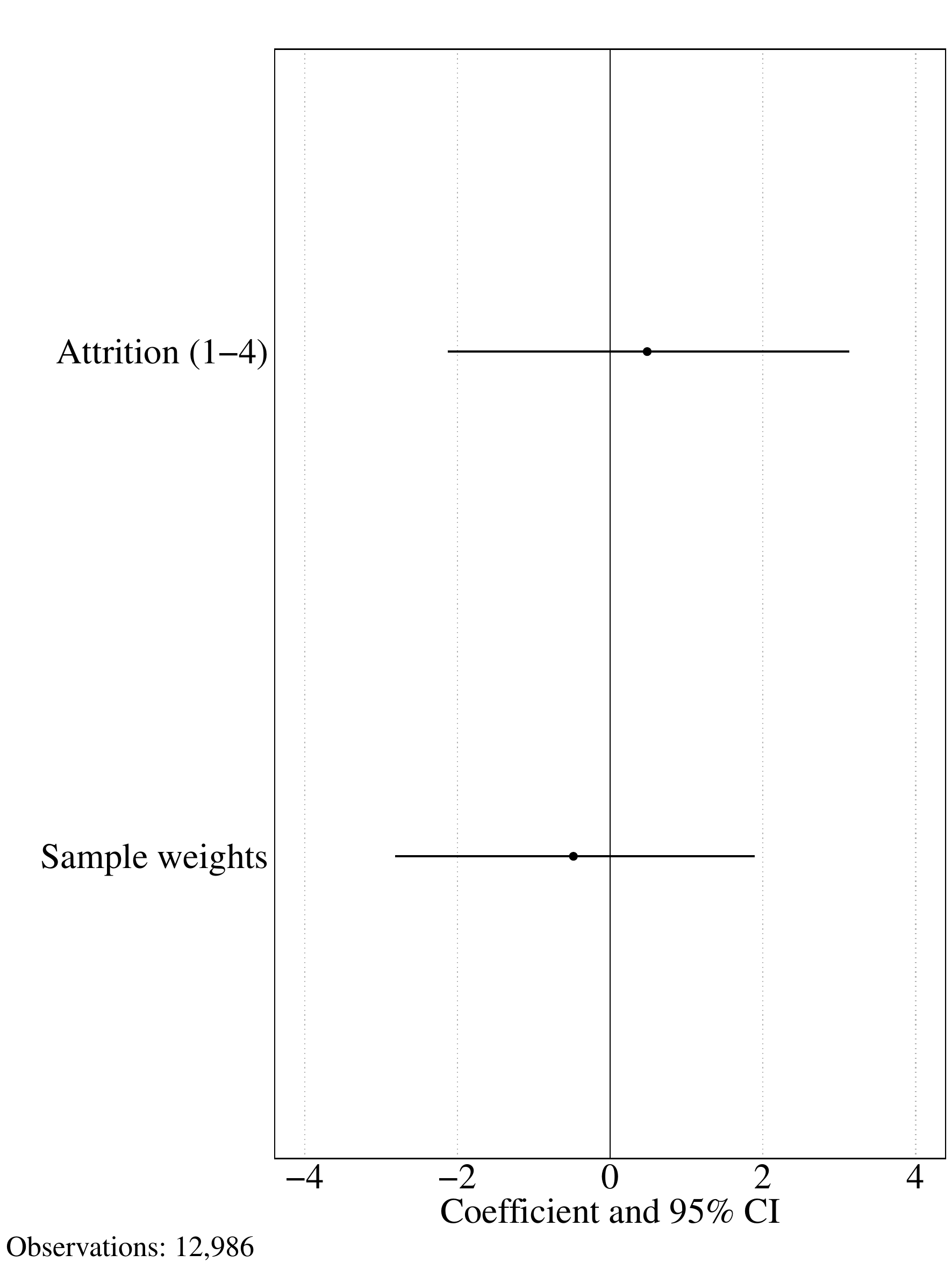}  
	\caption{Effects of the NHS on Selective Participation}
	\label{fig:participation_sample}
	\justifying 
	\footnotesize{Note: Each dot represents the causal effect of a separate regression on a different outcome. Standard errors under interference. The bandwidth is defined as 12 months pre- and post-NHS implementation. Source: UKB.}
\end{figure}


\begin{figure}[H]
	\centering
\includegraphics[height=0.55\textheight]{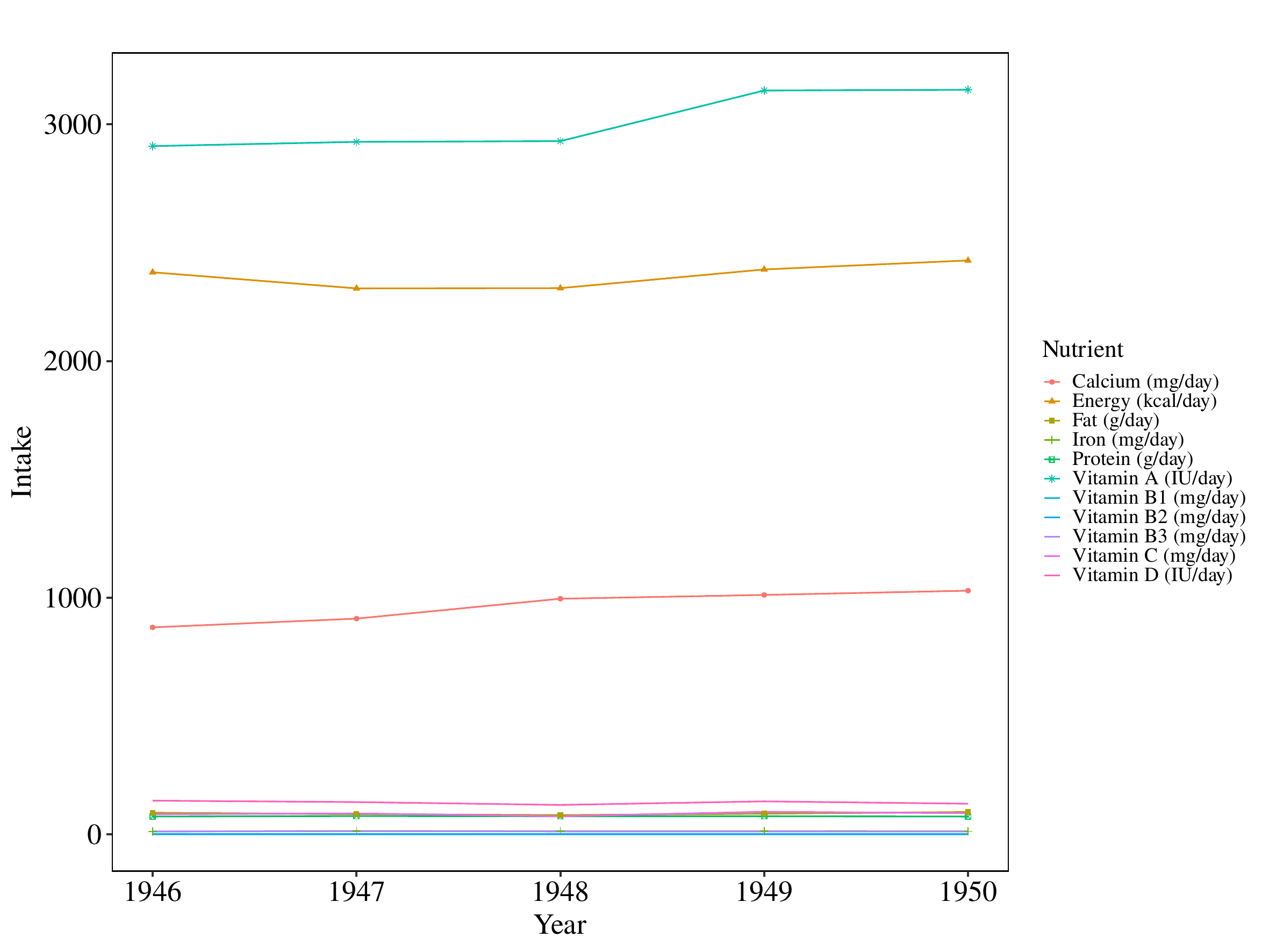}

	\caption{Average Daily Nutrient Intake Over Time}
	\label{fig:nutrients}
	\justifying
	\footnotesize{
	\textit{Notes:} The figure shows average daily nutrient intake per person among urban working-class households. 
	Energy is measured in kilocalories (kcal); protein and fat in grams; calcium, iron, and vitamins B$_1$, B$_2$, B$_3$, and C in milligrams; and vitamins A and D in International Units (IU). 
	Values are reported annually.
	\textit{Source:} National Food Survey.
	}
\end{figure}

 \begin{figure}[H]
	\centering
\includegraphics[height=0.55\textheight]{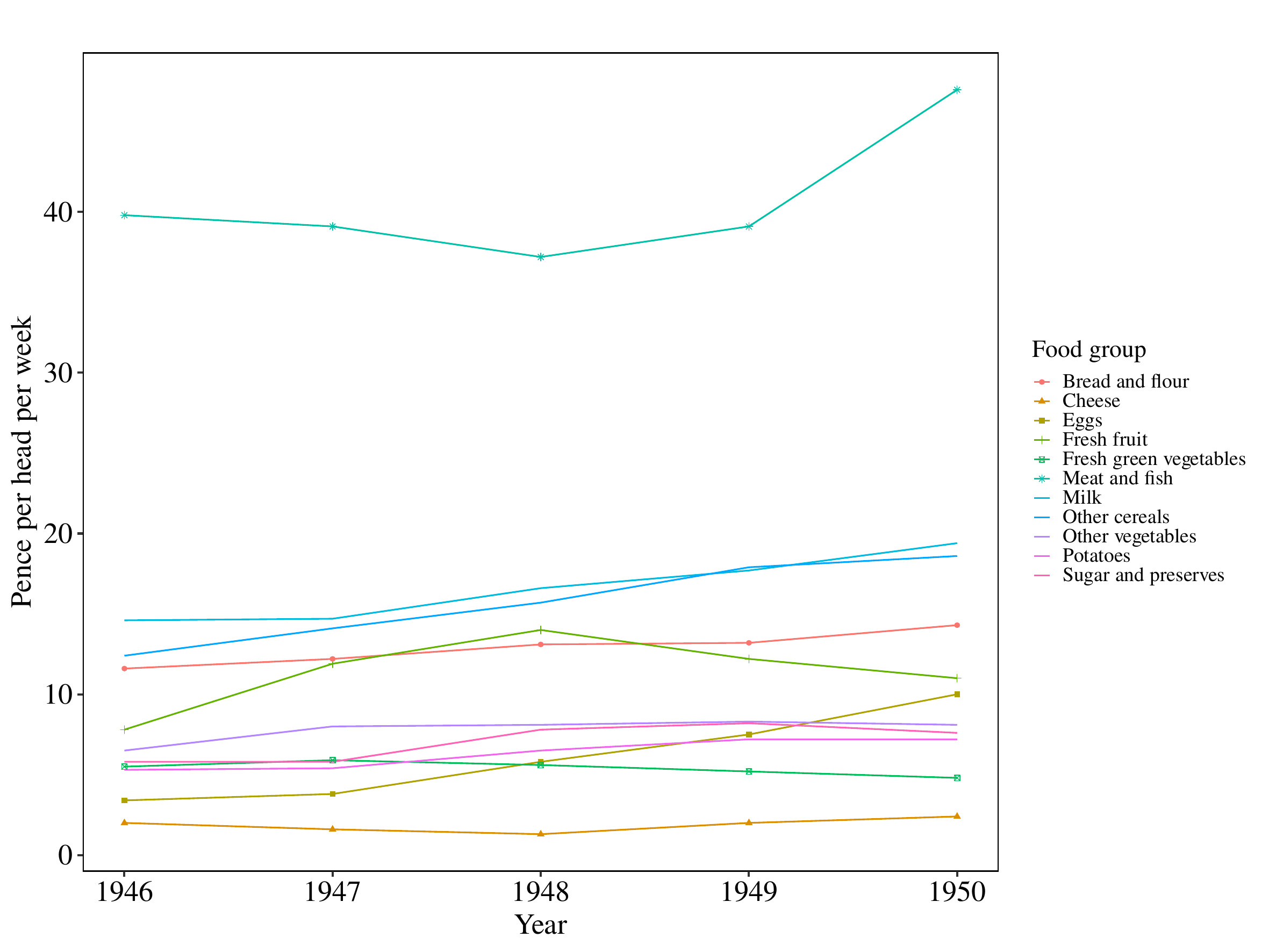}

	\caption{Average Weekly Expenditure Food Group Consumption Over Time}
	\label{fig:food_groups}
	\justifying
	\footnotesize{
	\textit{Notes:} The figure reports average weekly expenditure, in pence per head, for major food groups among urban working-class households. 
	Quantities are measured in grams per person per day and reported annually.
	\textit{Source:} National Food Survey.
	}
\end{figure}

		\begin{figure}[H]
			\centering
			\includegraphics[width=14cm]{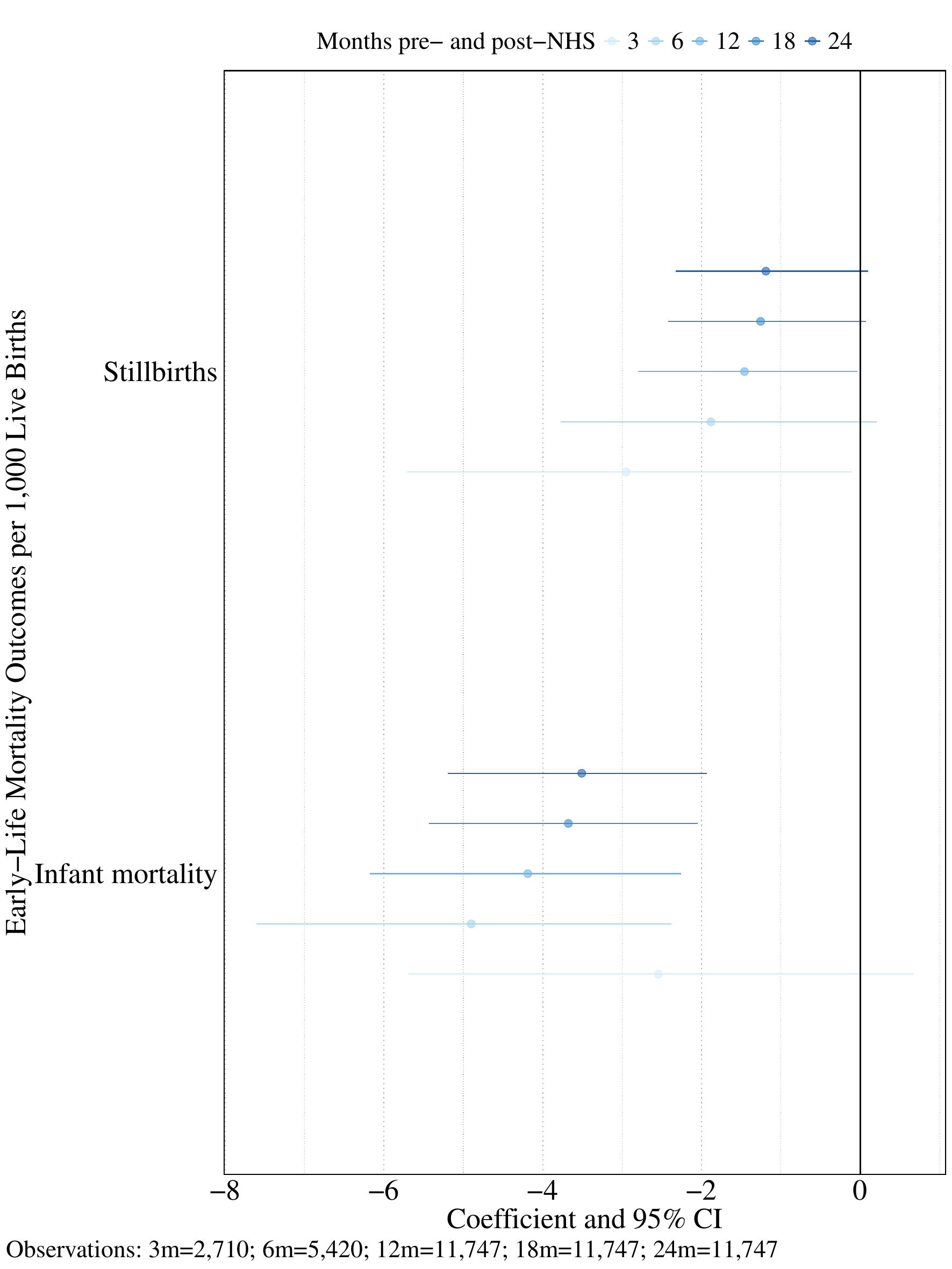}  
			\caption{ Effects of the NHS on Early-Life Mortality: Different Bandwidths}
			\label{fig:local_rand_windows_IMR}
			\justifying 
			\footnotesize{Note: RDD local randomization: confidence intervals use interference-robust standard errors. Source: Registrar General’s Weekly Returns.}
		\end{figure}

		\begin{figure}[H]
			\centering
			\includegraphics[width=14cm]{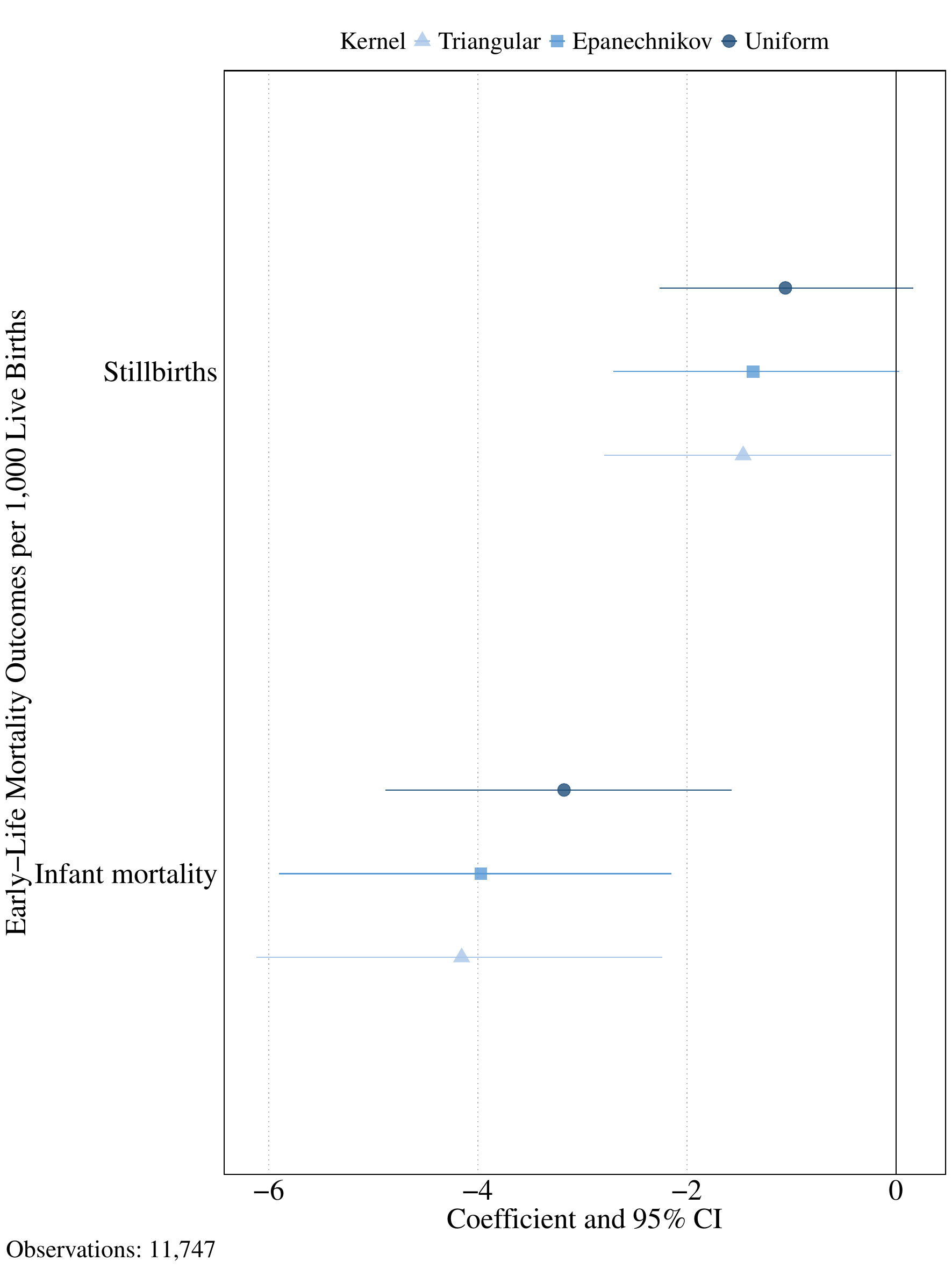}  
			\caption{Effects of the NHS on Early-Life Mortality: Different Kernels}
			\label{fig:IMR_kernel}
			\justifying 
			\footnotesize{Note: RDD local randomization: confidence intervals use interference-robust standard errors. Source: Registrar General’s Weekly Returns.}
		\end{figure}

		\begin{figure}[H]
			\centering
			\includegraphics[width=14cm]{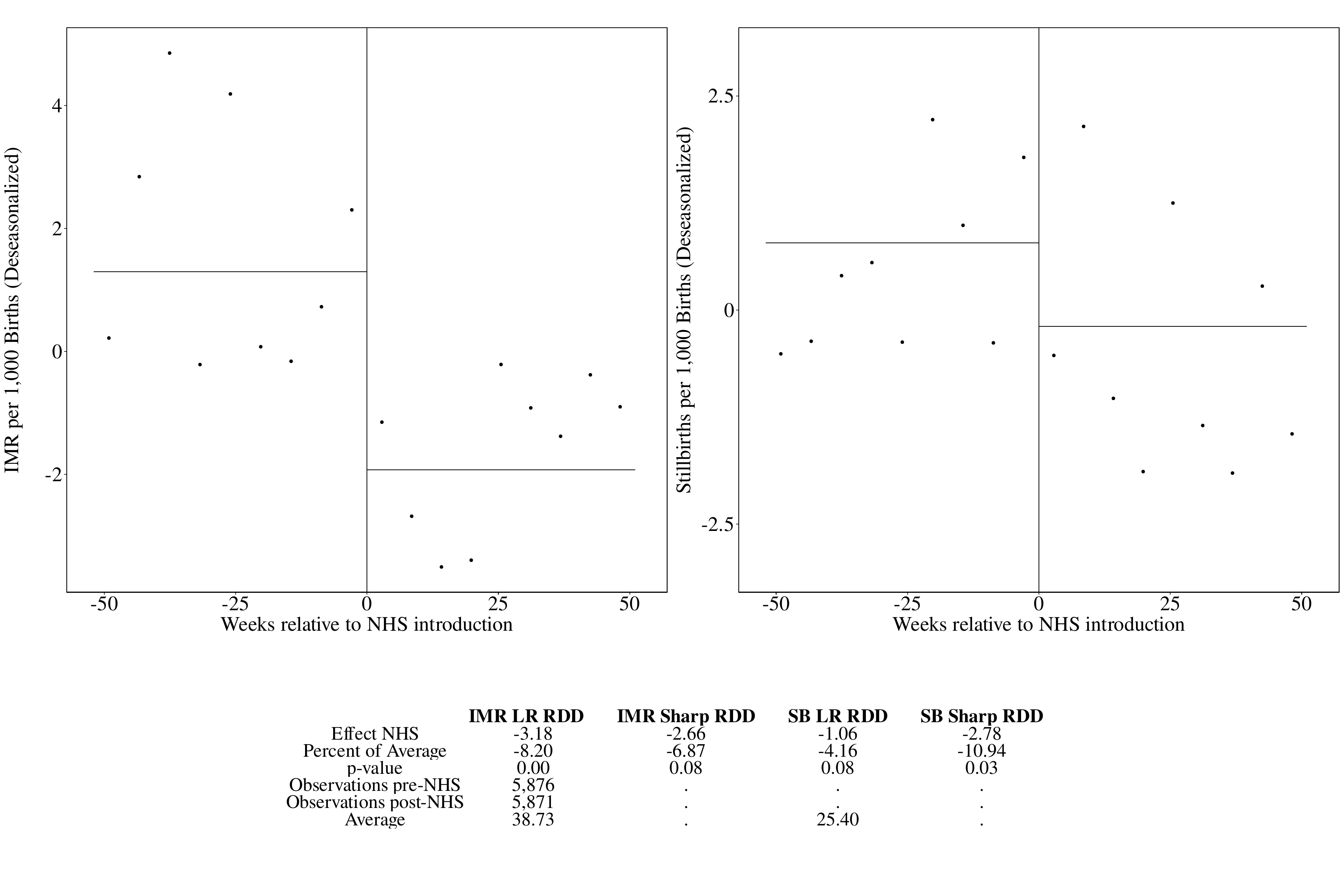}  
			\caption{Effects of the NHS on Early-Life Mortality: Doughnut Specification}
			\label{fig:mortality_weekly_donut_IMR_SB_only}
			\justifying 
			\footnotesize{Note: RDD local randomization: p-values are obtained from 1,000 permutations. Sharp RDD: conventional Eicker–Huber–White heteroskedasticity-robust standard errors. Source: Registrar General’s Weekly Returns.}
		\end{figure}

		\begin{figure}[H]
			\centering
			\includegraphics[width=14cm]{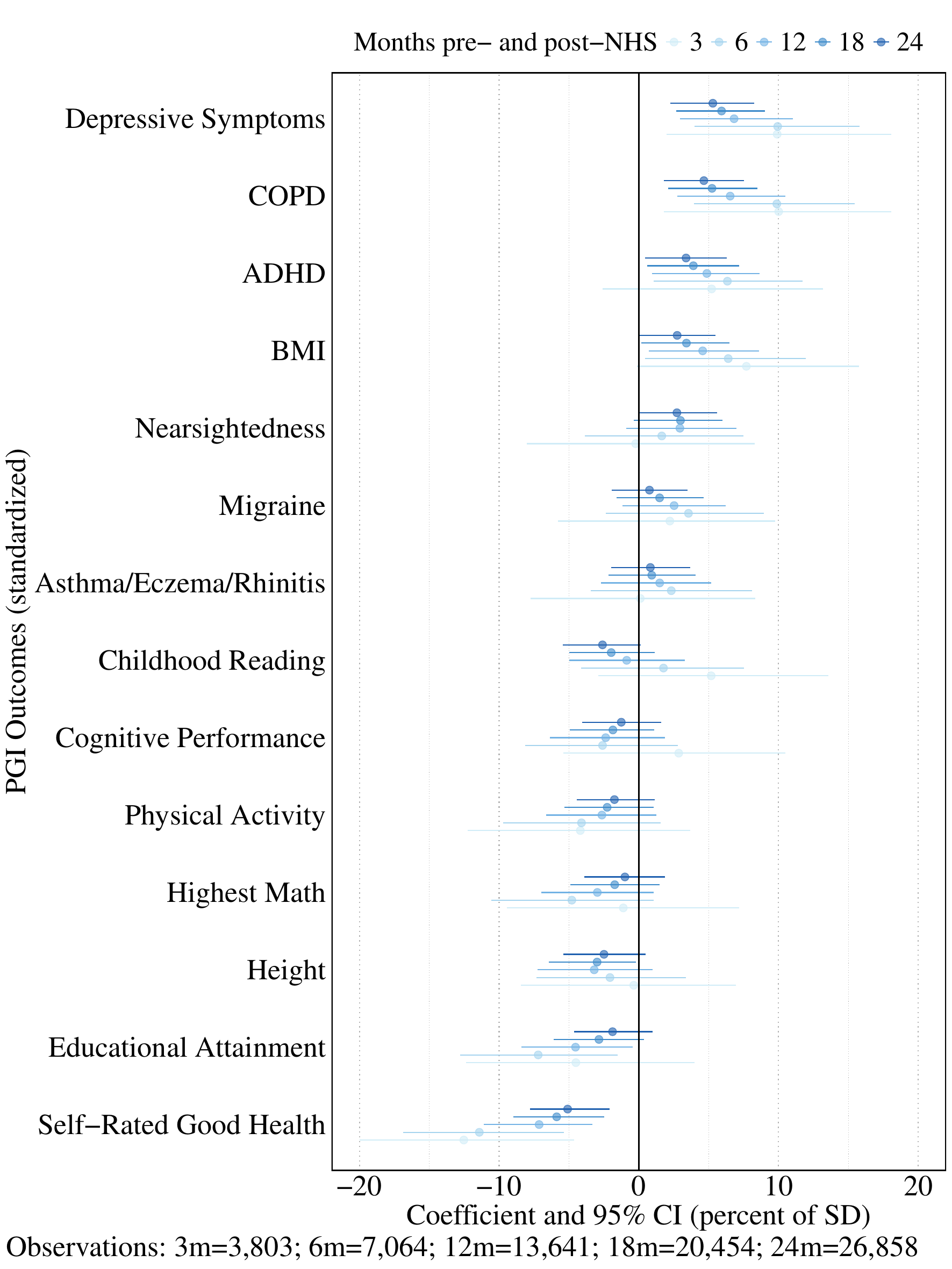}  
			\caption{ Effects of the NHS on PGIs: Different Bandwidths}
			\label{fig:wind}
			\justifying 
			\footnotesize{Note: Each dot represents the causal effect of a separate regression on a different outcome. Standard errors under interference. Source: UKB.}
		\end{figure}

		\begin{figure}[H]
			\centering
			\includegraphics[width=14cm]{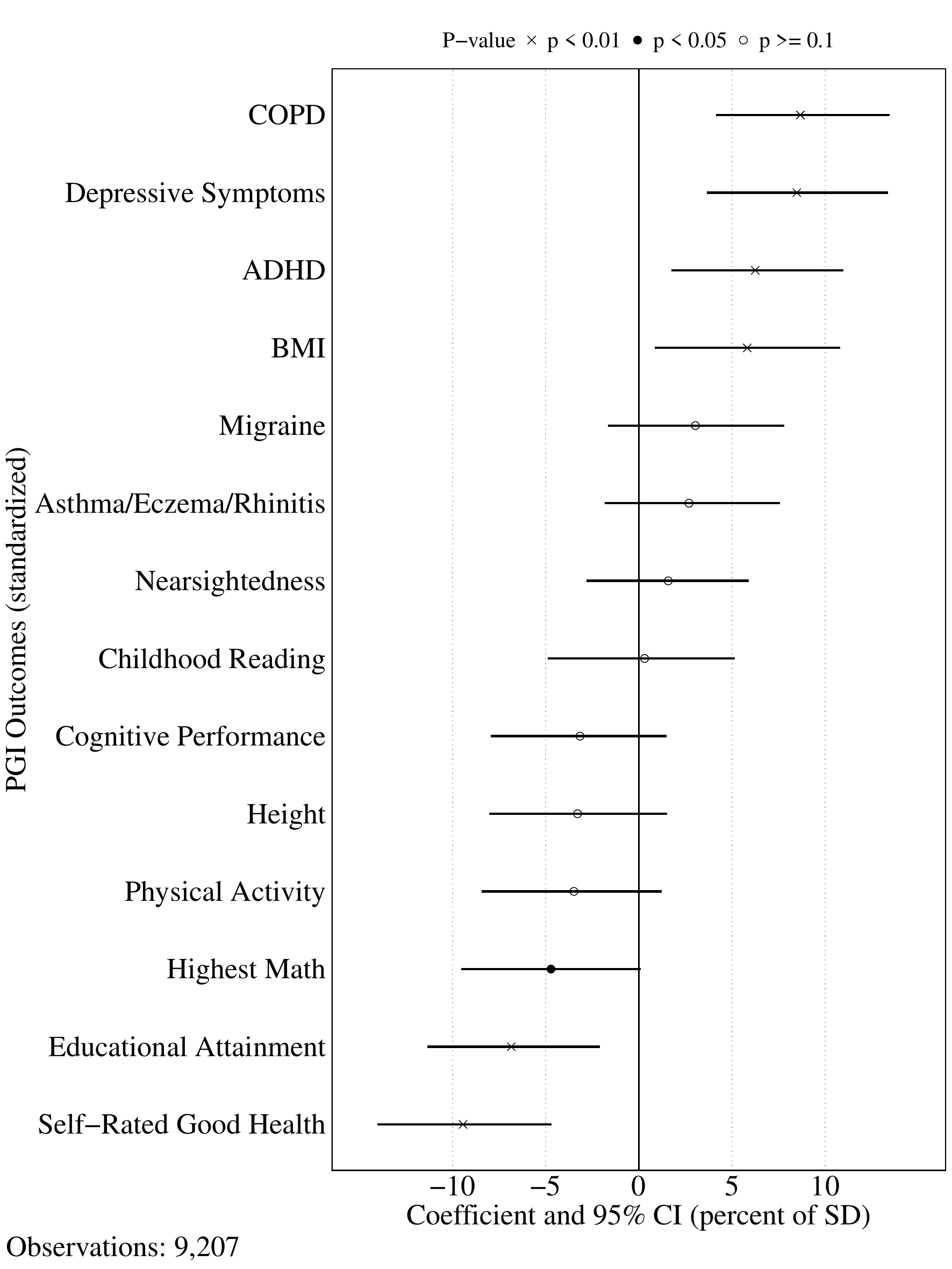}  
			\caption{ Effects of the NHS on PGIs: Balanced Bandwidths}
			\label{fig:fig2_bal}
			\justifying 
			\footnotesize{Note: Each dot represents the causal effect of a separate regression on a different outcome. P-values under permutation (1,000 permutations) and standard errors under interference. Source: UKB.}
		\end{figure}        

		\begin{figure}[H]
	\centering
		\includegraphics[width=14cm]{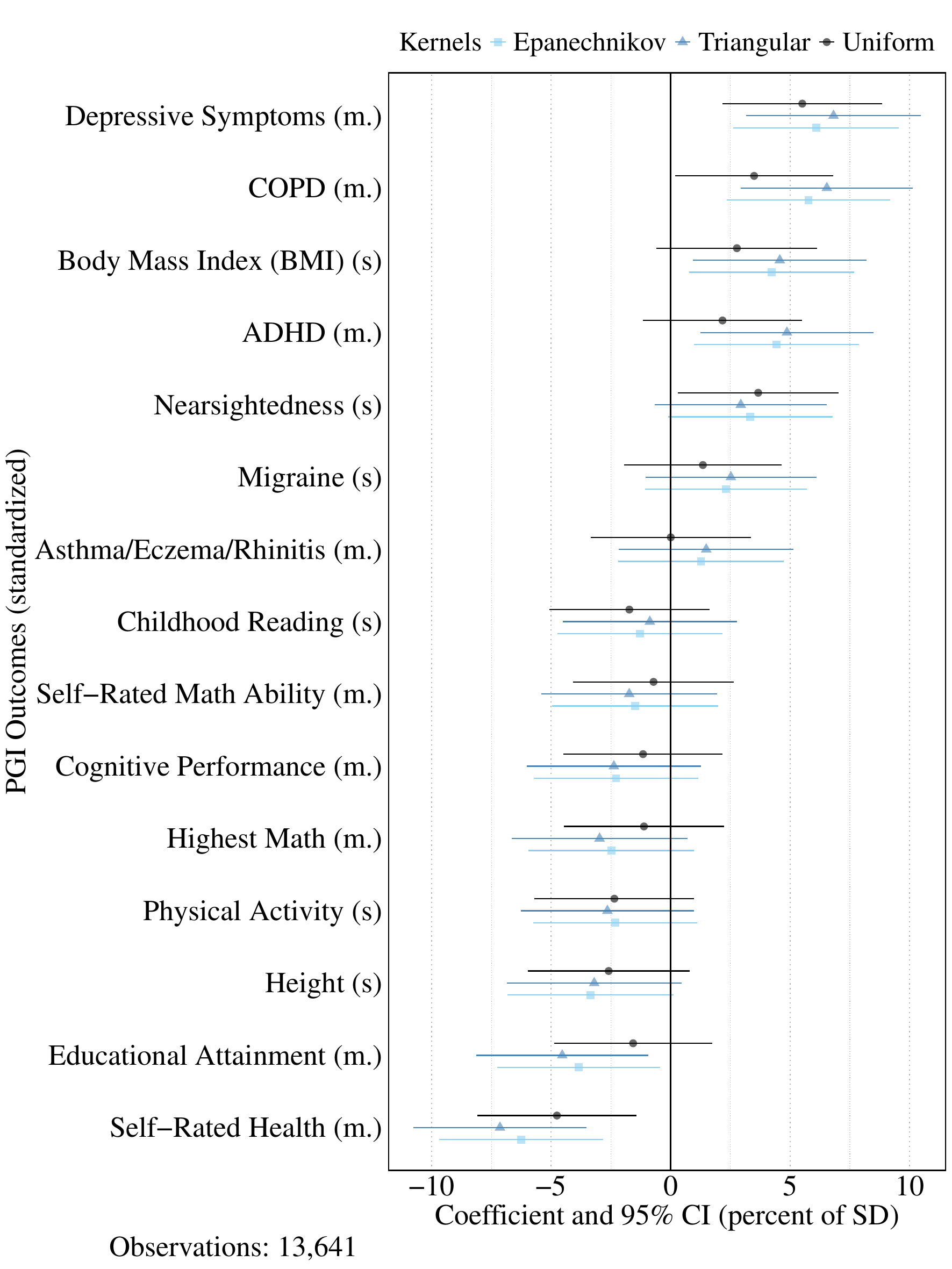}
	\caption{ Effects of the NHS on PGIs: Different Kernels}
	\label{fig:kern}
	\justifying 
	\footnotesize{Note: Each dot represents the causal effect of a separate regression on a different outcome. Standard errors under interference.  The bandwidth is 12 months pre- and post-NHS implementation. Source: UKB.}
\end{figure}

		\begin{figure}[H]
	\centering
	\includegraphics[width=14cm]{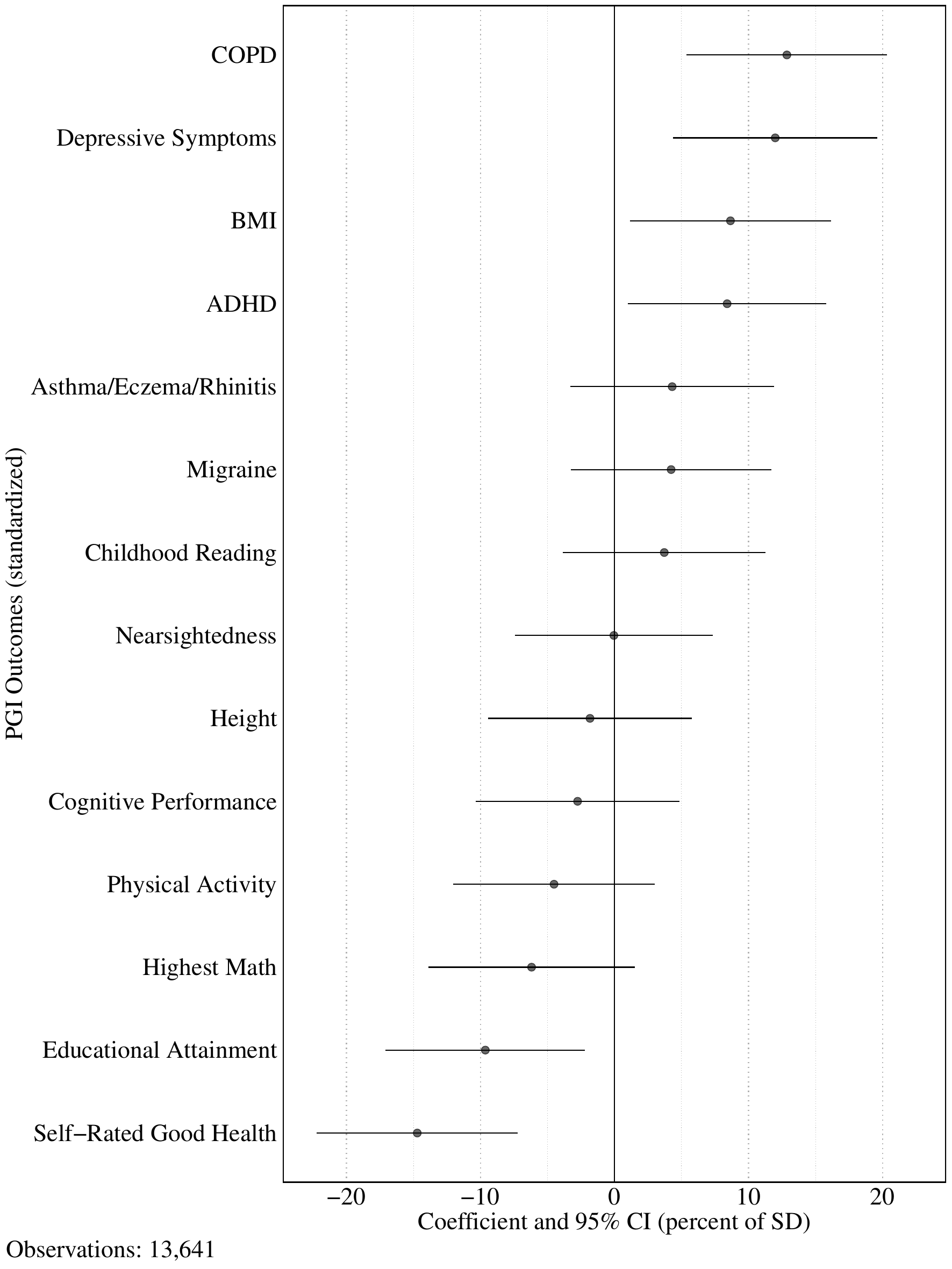}
	\caption{Effects of the NHS on PGIs: Using Sharp RDD}
	\label{fig:sharp}
	\justifying 
	\footnotesize{Note: Each dot represents the causal effect of a separate regression on a different outcome. 95\% confidence intervals. Cluster-robust variance estimation with degrees-of-freedom weights, combined with nearest neighbor variance estimation standard errors. The bandwidth is 12 months per- and post-NHS implementation. Linear fit with triangular kernels. Controls include age, gender, quarter of birth fixed effects, and 10 ancestry specific PCs. Source: UKB.}
\end{figure}

\begin{figure}[H]
	\centering
	\includegraphics[width=14cm]{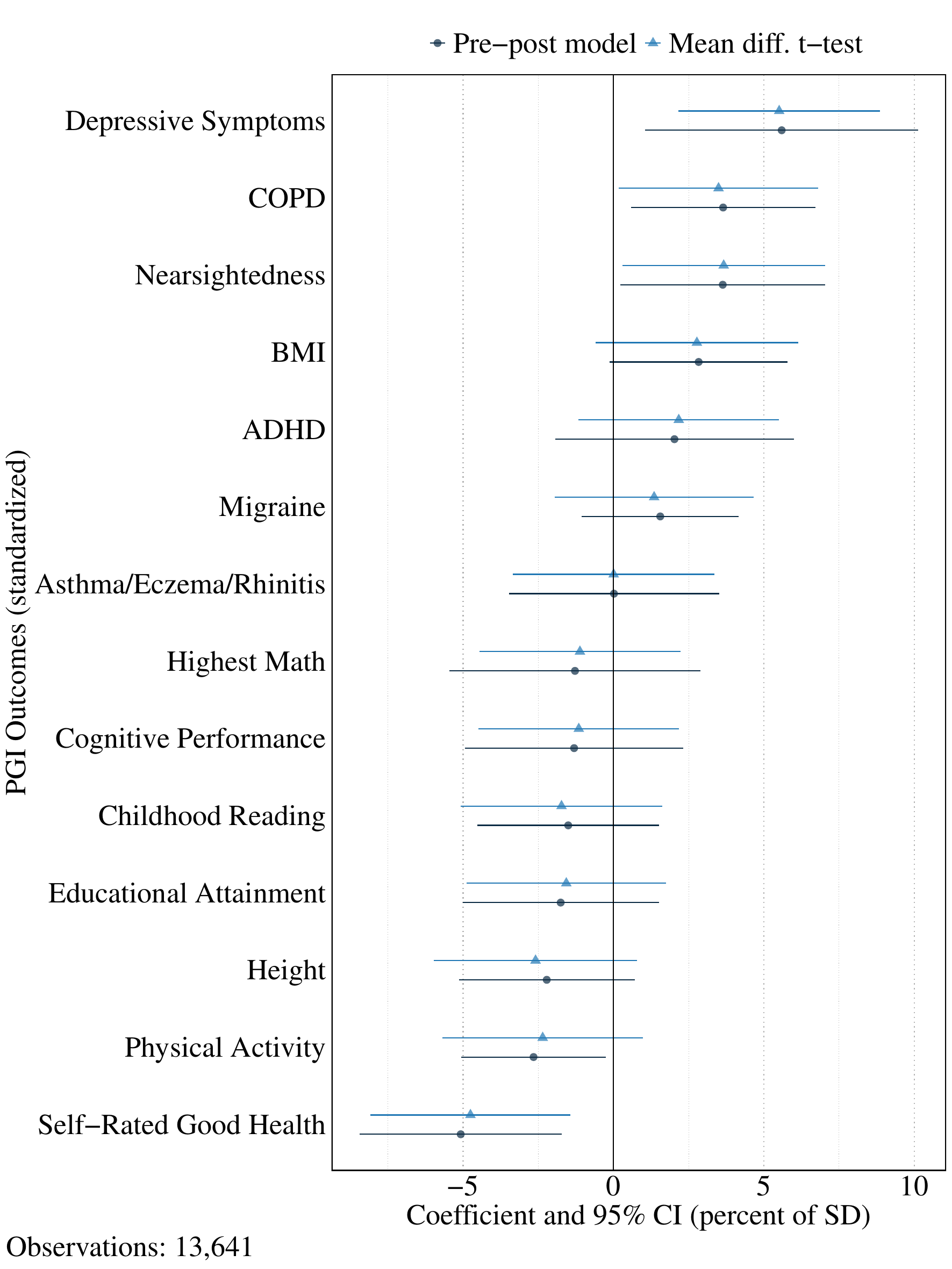}  
	\caption{Effects of the NHS on PGIs: Global Mean-comparison Approaches}
	\label{fig:pre_post}
	\justifying 
	\footnotesize{Note: Each dot represents the estimated causal effect from a separate regression on a different outcome. The Pre-post regression captures the effect of being born after the implementation of the NHS, controlling for sex, month and year of birth, interview year, and the first 10 genetic principal components (PCs). The Mean difference refers to the difference in average outcomes between pre- and post-NHS births, estimated using a two-sample t-test. Estimates are based on an 12-month bandwidth around the NHS implementation date. Source: UKB}
\end{figure}

\begin{figure}[H]
	\centering
	\includegraphics[width=14cm]{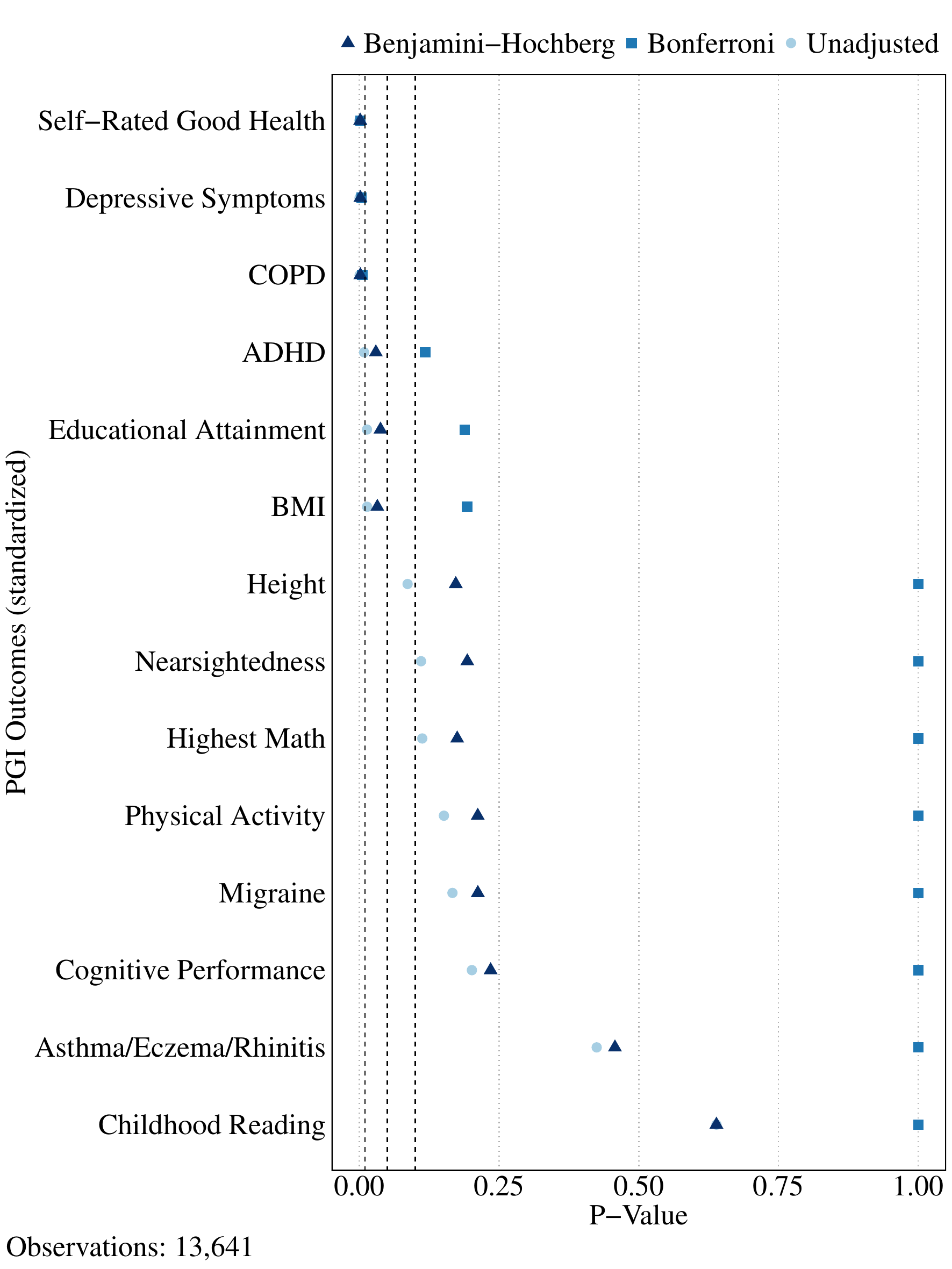}
	\caption{ Effects of the NHS on PGIs: Multiple Hypothesis Testing}
	\label{fig:mult}
	\justifying 
	\footnotesize{Note: The bandwidth is 12 months pre- and post-NHS implementation. Source: UKB.}
\end{figure}

\begin{figure}[H]
	\centering
	\includegraphics[height=0.7\textheight]{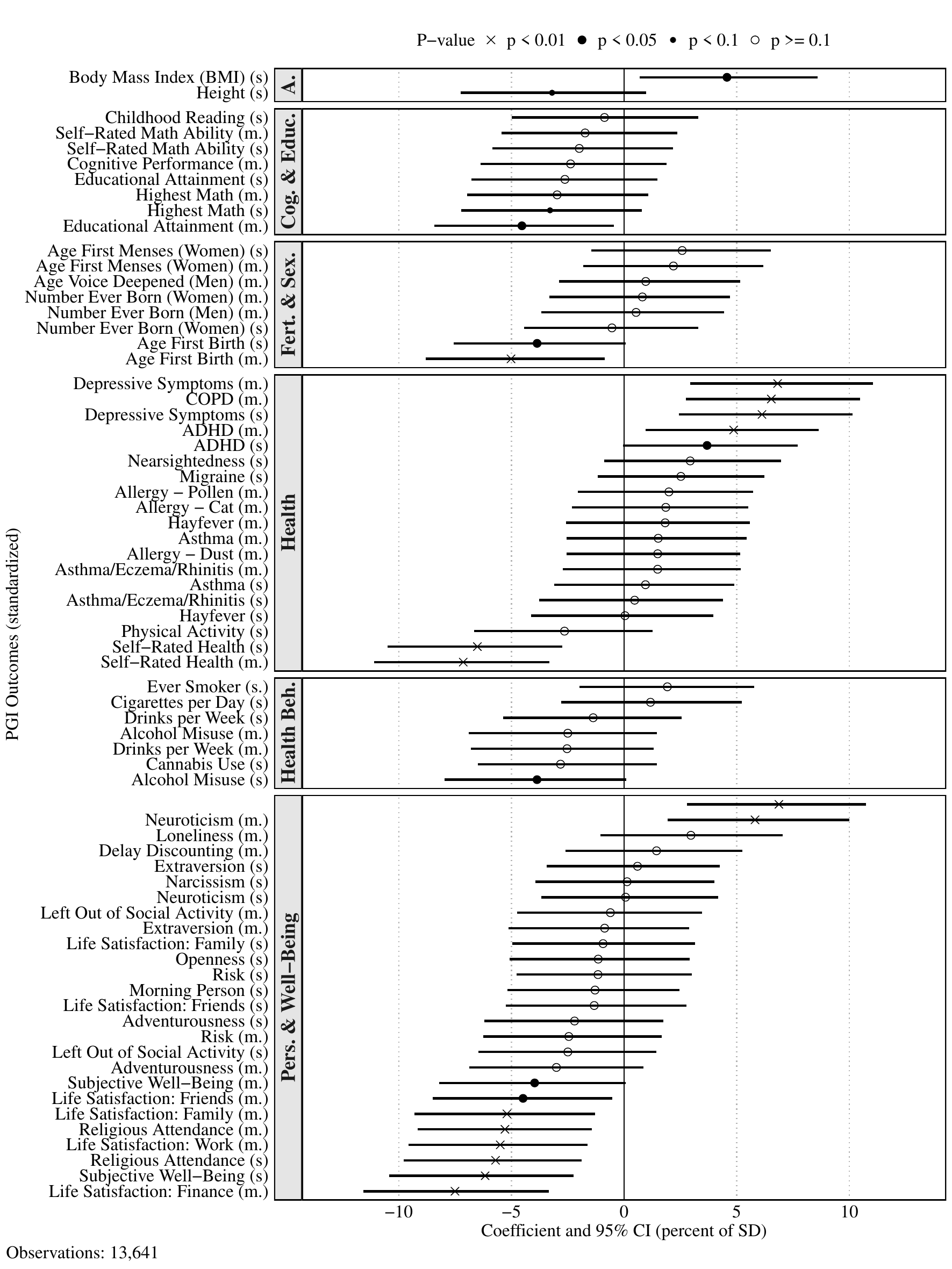}  
	\caption{ Effects of the NHS on all PGIs}
	\label{fig:allPGI}
	\justifying 
	\footnotesize{Note: Each dot represents the causal effect of a separate regression on a different outcome. P-values under permutation (1,000 permutations) and standard errors under interference. The bandwidth is 12 months pre- and post-NHS implementation. Source: UKB.}
\end{figure}

\begin{figure}[ht]
	\centering
\includegraphics[width=12cm]{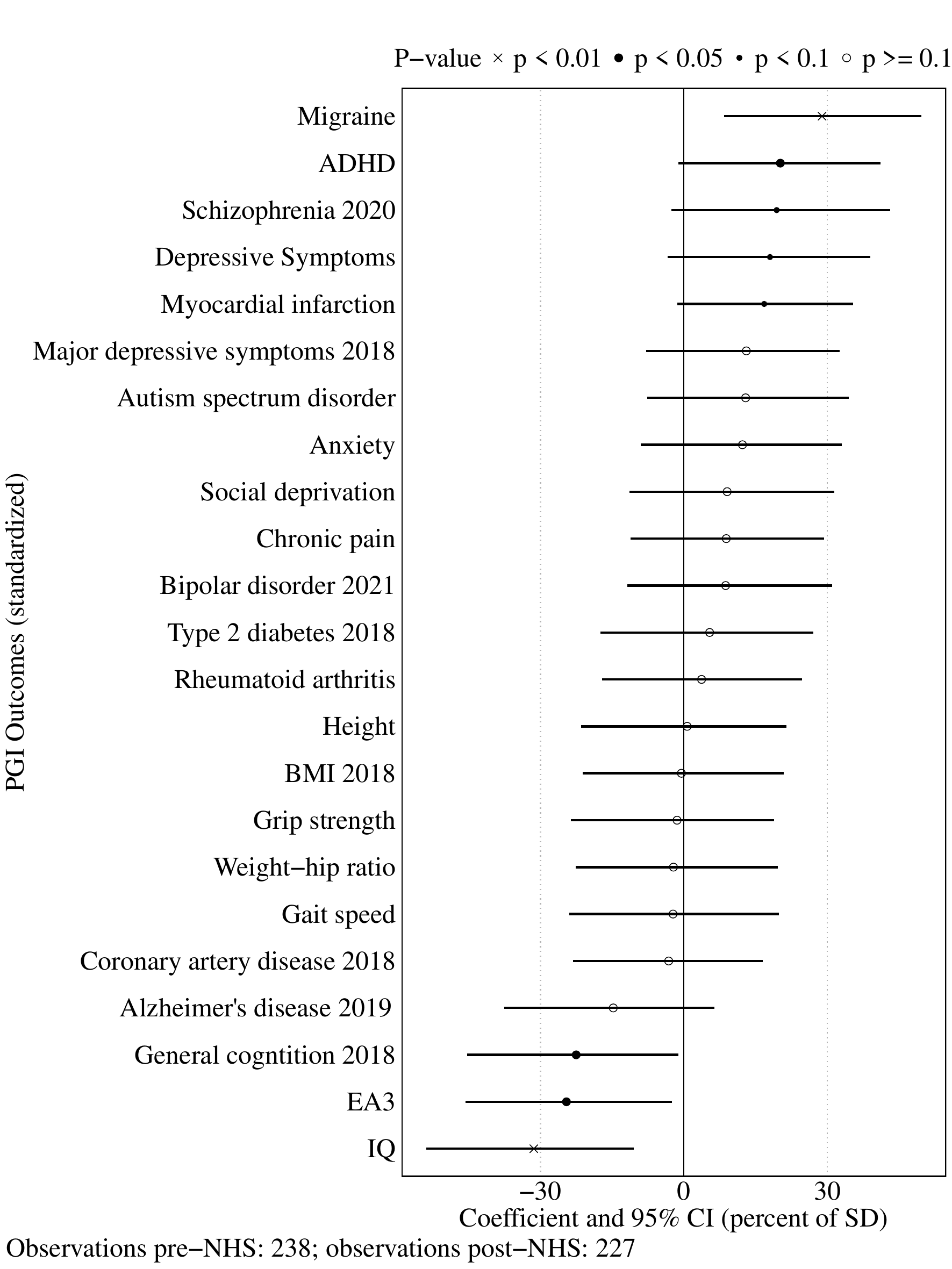}
	\caption{Effects of the NHS on PGIs: Using ELSA}
	\label{fig:elsa}
	\justifying 
	\footnotesize{Note: Each dot represents the causal effect of a separate regression on a different outcome. P-values under permutation (1,000 permutations) and standard errors under interference. The bandwidth is 12 months pre- and post-NHS implementation. Source: UKB.}
\end{figure}

\begin{figure}[ht]
	\centering
\includegraphics[width=12cm]{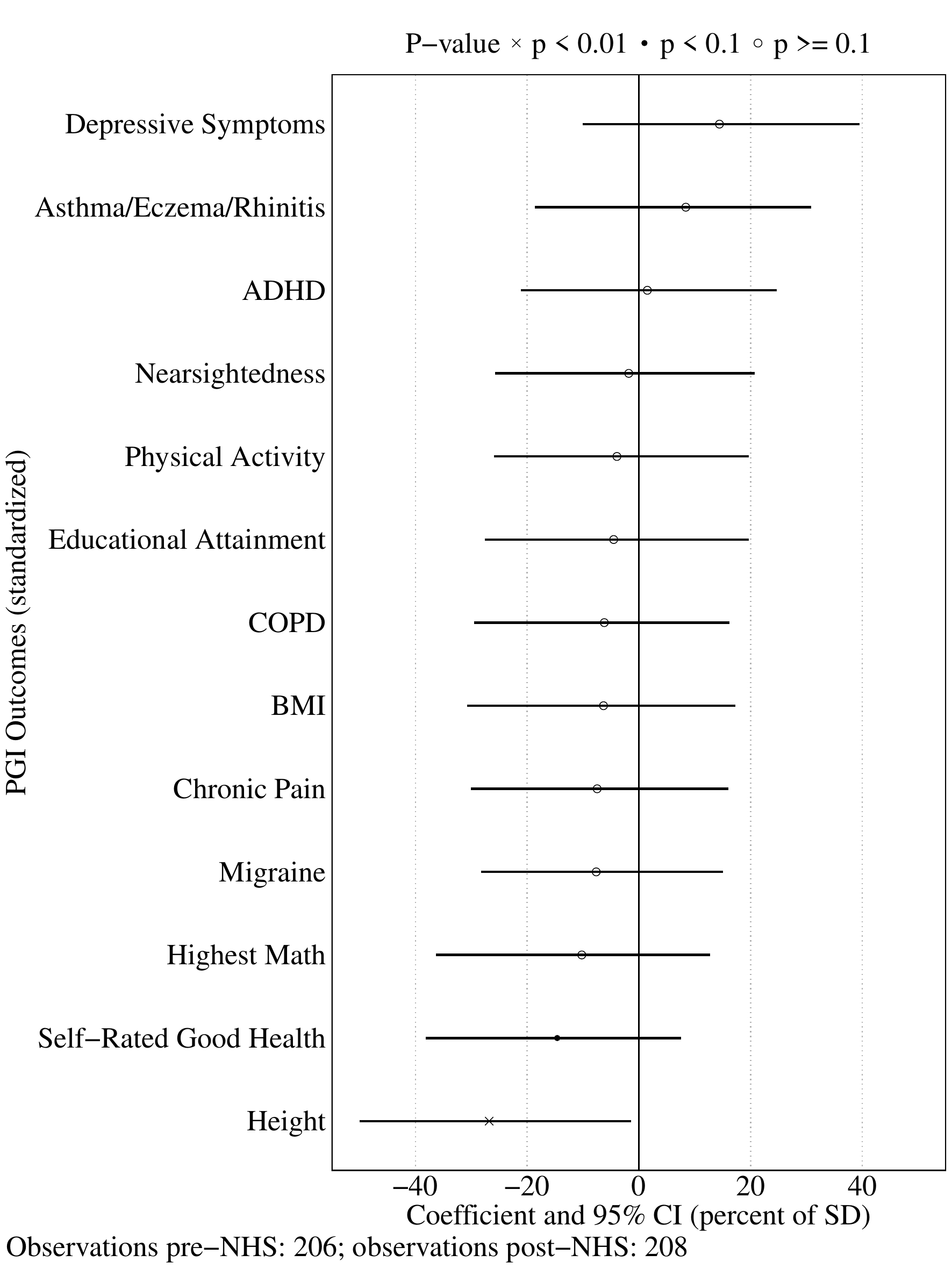}
	\caption{Effects of the NHS on PGIs: Using USoc}
	\label{fig:USoc}
	\justifying 
	\footnotesize{Note: Each dot represents the causal effect of a separate regression on a different outcome. P-values under permutation (1,000 permutations) and standard errors under interference. The bandwidth is 12 months pre- and post-NHS implementation. Source: UKB.}
\end{figure}

\begin{figure}[ht]
	\centering
\includegraphics[width=12cm]{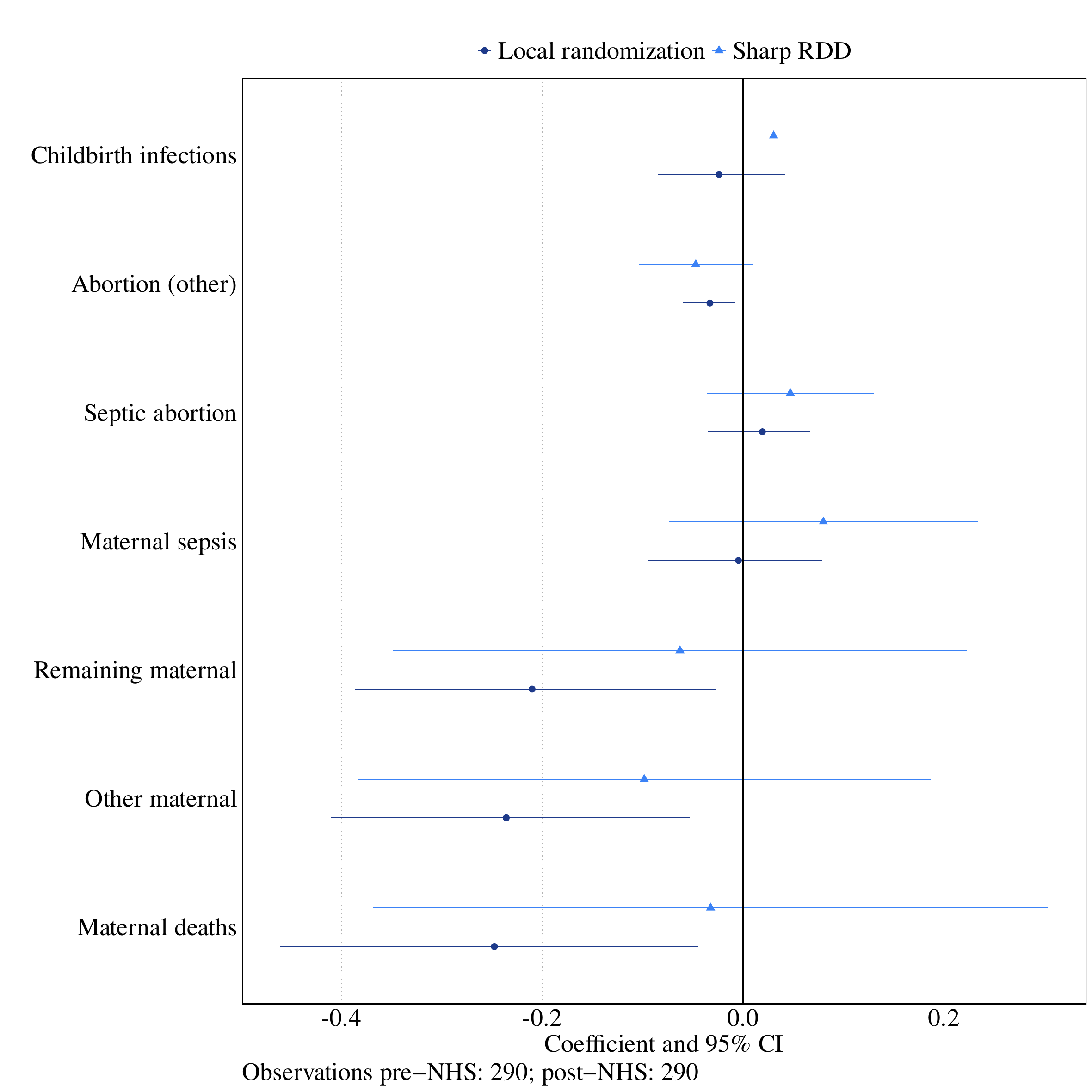}
	\caption{Effects of the NHS on Maternal Mortality}
	\label{fig:maternal}Effects of the NHS on Fertility
	\justifying 
	\footnotesize{Note: Each dot represents the causal effect of a separate regression on a different outcome. RDD local randomization: confidence intervals use interference-robust standard errors. Sharp RDD: conventional Eicker–Huber–White heteroskedasticity-robust standard errors. The bandwidth is two years pre- and post-NHS implementation. Source: Registrar General’s Weekly Returns.}
\end{figure}

\begin{figure}[ht]
	\centering
\includegraphics[width=12cm]{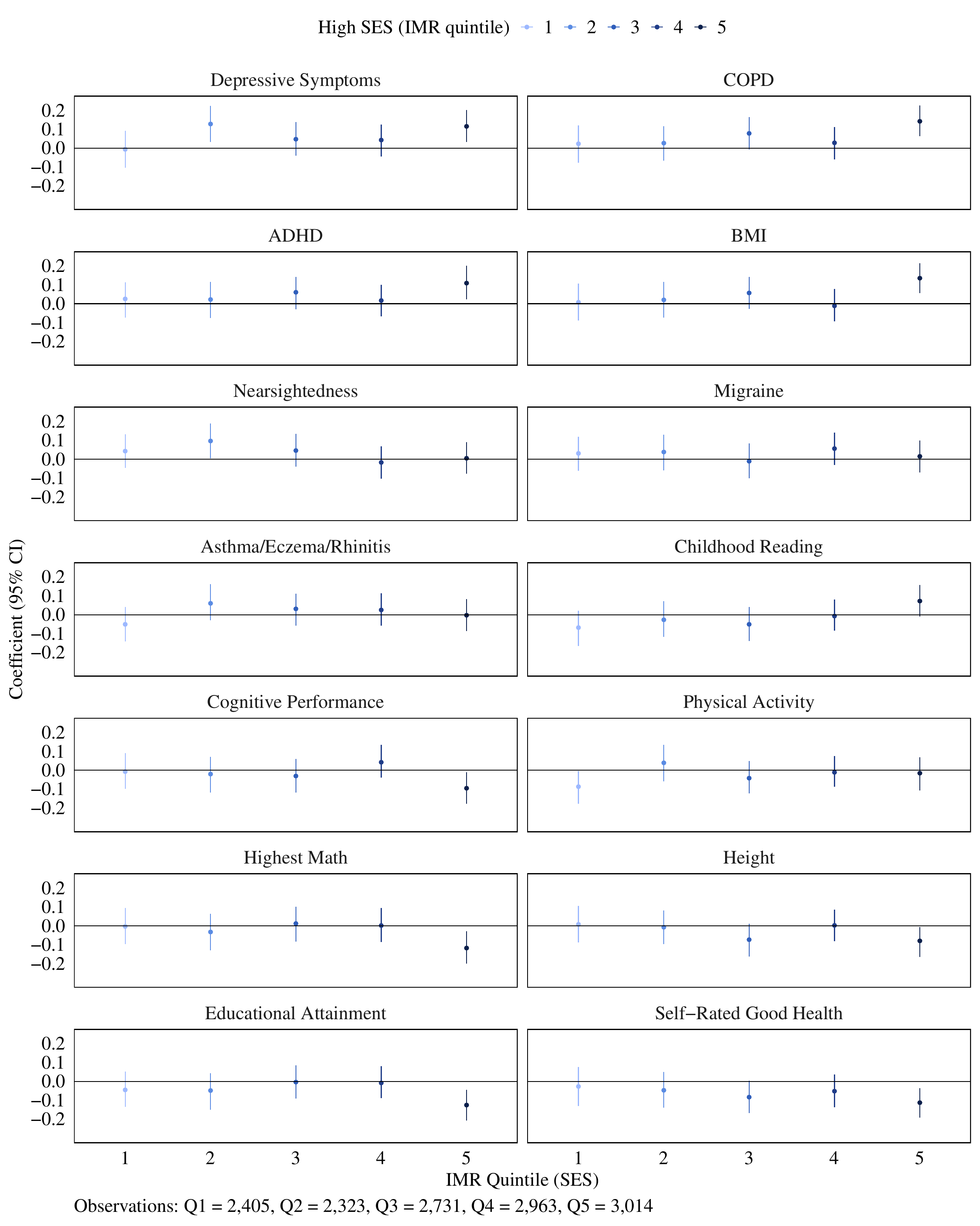}
	\caption{Effects of the NHS on PGIs: Heterogeneity by pre-NHS IMR}
	\label{fig:heterogeneity_all}
	\justifying 
	\footnotesize{Note: Each dot represents the causal effect of a separate regression on a different outcome. Standard errors under interference. The bandwidth is 12 months pre- and post-NHS implementation. Source: UKB and Registrar General’s Statistical Review.}
\end{figure}

\begin{figure}[ht]
	\centering
\includegraphics[width=12cm]{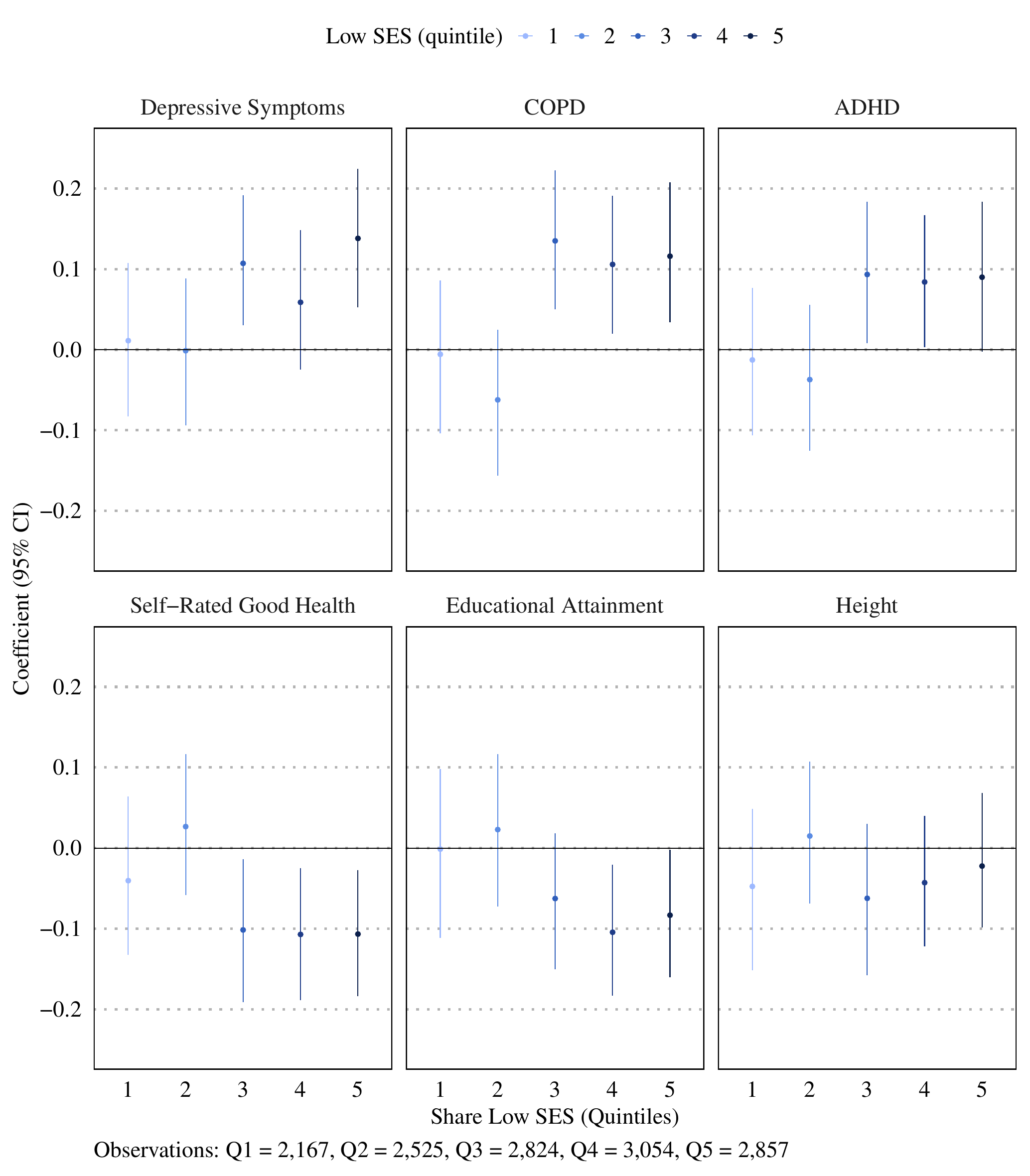}
	\caption{Effects of the NHS on PGIs: Heterogeneity by Low SES Quintiles (1951 Census)}
	\label{fig:heterogeneity_census}
	\justifying 
	\footnotesize{Note: Each dot represents the causal effect of a separate regression on a different outcome. Standard errors under interference. The bandwidth is 12 months pre- and post-NHS implementation. Source: UKB and Census 1951.}
\end{figure}

\end{appendices}

\end{document}